\documentclass[%
 reprint,
 twocolumn,
 table,
 longbibliography,
 amsmath,amssymb,
 aps,
]{revtex4-1}

\usepackage{enumitem}
\usepackage[table,xcdraw]{xcolor}
\usepackage{graphicx}
\usepackage{bm}

\usepackage{amsmath}
\usepackage{booktabs}
\usepackage{color}
\usepackage{orcidlink}
\usepackage{hyperref}
\usepackage{soul}
\hypersetup{
  colorlinks=true,
  urlcolor=blue,
  linkcolor=red,
  citecolor=blue
}

\usepackage{multirow}
\newcolumntype{C}[1]{>{\centering\arraybackslash}p{#1}}

\DeclareFontFamily{U}{mathb}{\hyphenchar\font45}
\DeclareFontShape{U}{mathb}{m}{n}{
      <5> <6> <7> <8> <9> <10> gen * mathb
      <10.95> mathb10 <12> <14.4> <17.28> <20.74> <24.88> mathb12
      }{}
\DeclareSymbolFont{mathb}{U}{mathb}{m}{n}
\DeclareFontSubstitution{U}{mathb}{m}{n}

\makeatletter
\newcommand\RedeclareMathOperator{%
  \@ifstar{\def\rmo@s{m}\rmo@redeclare}{\def\rmo@s{o}\rmo@redeclare}%
}
\newcommand\rmo@redeclare[2]{%
  \begingroup \escapechar\m@ne\xdef\@gtempa{{\string#1}}\endgroup
  \expandafter\@ifundefined\@gtempa
     {\@latex@error{\noexpand#1undefined}\@ehc}%
     \relax
  \expandafter\rmo@declmathop\rmo@s{#1}{#2}}
\newcommand\rmo@declmathop[3]{%
  \DeclareRobustCommand{#2}{\qopname\newmcodes@#1{#3}}%
}
\@onlypreamble\RedeclareMathOperator
\DeclareMathSymbol{\curvearrowtopleft} {\mathrel}{mathb}{"F0}
\DeclareMathSymbol{\curvearrowbotright}{\mathrel}{mathb}{"F4}
\DeclareMathSymbol{\curvearrowtopright}{\mathrel}{mathb}{"F1}

\makeatother

\newcommand{\ee}{\end{equation}}
\newcommand{\be}{\begin{equation}}
\newcommand{\bea}{\begin{eqnarray*}}
\newcommand{\eea}{\end{eqnarray*}}
\newcommand{\bean}{\begin{eqnarray}}
\newcommand{\eean}{\end{eqnarray}}
\newcommand{\ba}{\begin{array}}
\newcommand{\ea}{\end{array}}

\newcommand{\cas}{\textstyle\frac}
\newcommand{\red}{}
\newcommand{\blue}{}

\newcommand{\e}{{\mathrm{e}}}
\renewcommand{\i}{{\mathrm{i}}}
\renewcommand{\d}{{\mathrm{d}}}

\newcommand{\cgr}{\cellcolor{gray!20}}

\setlist[itemize]{leftmargin=0pt, itemindent=*, labelsep=6pt, topsep=0pt, itemsep=2pt, parsep=0pt}

\defcitealias{PS_22}{PS22}

\begin{document}

\preprint{APS/123-QED}

\title{Landau damping of disturbances in nearly inviscid inflectional shear flows}

\author{Evgeny V. Polyachenko}
\affiliation{Institute of Astronomy, Russian Academy of Sciences, 
48 Pyatnitskaya st, Moscow 119017, Russia}
\email{epolyach@inasan.ru}
\affiliation{SnT SEDAN, University of Luxembourg, 
29 boulevard JF Kennedy, L-1855 Luxembourg, Luxembourg}
\email{evgeny.polyachenko@uni.lu}

\author{Ilia G. Shukhman}
\affiliation{Institute of Solar-Terrestrial Physics, Russian Academy of Sciences, \\ Siberian Branch, P.O. Box 291, Irkutsk 664033, Russia}
\email{shukhman@iszf.irk.ru}

\author{Michael Karp}
\affiliation{The Stephen B. Klein Faculty of Aerospace Engineering, Technion Israel Institute of Technology, Haifa 3200003, Israel}
\email{mkarp@technion.ac.il}

\date{\today}

\begin{abstract}
We investigate the structure of damped two-dimensional perturbations in unstable plane-parallel shear flows with an inflection point. In inviscid flows within the stable wavenumber region $k$, no regular eigenmodes exist -- the frequency spectrum $\omega$ consists of a continuous set of singular van Kampen modes with real frequencies. Nevertheless, initial perturbations of the total vorticity integrated across the flow decay exponentially, resembling the behavior of an eigenmode with complex eigenfrequency ${\rm Im}\,\omega<0$ (Landau damping). However, the vorticity itself does not decay but becomes increasingly corrugated across the flow. We demonstrate that accounting for arbitrarily small viscosity transforms this exponentially decaying perturbation into a true eigenmode in which the vorticity preserves its spatial form. We numerically trace the transformation of the vorticity structure of this mode and its disappearance as viscosity approaches zero. We discuss similarities and differences between the behavior of damped perturbations in the transition from inviscid to nearly inviscid flows in hydrodynamics and their behavior in plasma and homogeneous stellar systems during the analogous transition from collisionless to very weakly collisional systems.
\end{abstract}

\pacs{}

\maketitle

\section{Introduction} \label{sec:intro}

For inviscid plane-parallel shear flows $V_x=U(y)$ with an inflection point, the eigenmode spectrum \red{of 2D perturbations $ \sim \exp[\i (kx-\omega t)]$} in the stable wavenumber region $k>k_{\rm crit}$ forms a continuous set of van Kampen modes with real frequencies \cite{vK_55, Case59}. However, when we consider the initial value problem, the total vorticity across the flow (see Sec.\,\ref{sec:evo} for the definition of total vorticity) exhibits prolonged exponential decay under certain conditions, followed by slower algebraic decay. This exponential decay constitutes Landau damping \cite{Lan_46}, representing a superposition of singular van Kampen modes rather than true eigenmodes. The vorticity itself becomes increasingly corrugated over time. Such perturbations are termed Landau {\it quasi-modes} (see, e.g., \cite{PS_22, Balmforth_Morrison}).

\blue{Key insights into Landau damping emerged from the powerful analogy connecting wave-particle resonance in plasmas to critical layers in inviscid shear flows. Briggs {\it{et al.}}~\cite{BDL_1970} theoretically investigated the role of Landau damping in both electron beams and inviscid shear flows. Later, Corngold~\cite{Corngold_1995} developed analytical results for specific density profiles, which were later generalized into a numerical scheme by Spencer \& Rasband~\cite{Spencer_Rasband_1997}. Furthermore, Schecter {\it et al.}~\cite{Schecter_2000} examined the inviscid damping of quasi-modes in a 2D vortex and related their results to an electron plasma experiment.}

The question arises: what happens to Landau quasi-modes when arbitrarily small viscosity is included? In electron plasma and stellar media, studies by Ng and co-authors \cite{Ng1999,Ng2004,Ng2021}, Auerbach \cite{Auerbach_77}, and Polyachenko and Shukhman \cite{PS_2025,PS_25PRE} showed that collisions transform Landau quasi-modes into true modes with strongly oscillatory structure in the resonance region. The eigenvalues receive only small corrections, essentially preserving the Landau eigenvalues.

For shear flows, the Landau quasi-mode behavior and their viscous transformation into true Landau modes differs significantly from that in electron plasma and stellar systems:

\begin{itemize}

\item \textbf{Asymptotic behavior:} 
{Unlike plasma and \blue{homogeneous} stellar media, exponential Landau damping of the total vorticity occurs only during an intermediate time period, with ultimate power-law decay taking over at late times.}

\item \textbf{Limited quasi-mode spectrum:} Inviscid shear flows possess only a finite number of known Landau quasi-modes, contrasting with countable sets in plasma and {homogeneous} stellar media. For example, for $U(y) = \tanh(y)$, only two aperiodic quasi-modes exist \cite{PS_22}. Similarly, for the stable inflectional flow $U(y) = y + \beta y^3$ in the channel, only a pair of oscillating quasi-modes exists \cite {PS_24}. This limited spectrum prevents the construction of complete eigenmode systems originating from Landau quasi-modes in weakly viscous flows, which are needed to trace arbitrary disturbance evolution, unlike the complete sets available in plasma \cite{Ng1999,Ng2004} and \blue{homogeneous} stellar media \cite{Ng2021}.

\item \textbf{Quasi-mode survival:} Not every Landau quasi-mode survives viscosity inclusion. Some of them completely disappear under arbitrarily small viscosity. For example, we verified that the quasi-modes found in \cite{PS_24} for the flow $U=y+\beta y^3$ disappear.

\item \textbf{Predictive limitations:} No general criterion exists to establish quasi-mode existence based solely on velocity profile characteristics. Nevertheless, unstable shear flows with inflection points  appear to possess at least one Landau quasi-mode in the stable wavenumber region that continuously transforms into a true unstable mode in the unstable wavenumber region (see, e.g., Fig.\,5 in \cite{PS_22}).

\end{itemize}

The question of whether disturbance evolution in high Reynolds number flows can be described by purely inviscid equations has long interested researchers (see, e.g., Lin's book \cite{Lin_55}). However, previous studies have focused primarily on the influence of viscosity on stability and transition, rather than on how viscosity affects the evolution of disturbances that decay even within the purely inviscid framework.

In this paper, we investigate the transformation of the least damped Landau quasi-mode into a true eigenmode under vanishingly small viscosity, using the free mixing layer $U(y)=\tanh(y)$ as an example. We track how both the eigenfrequency (phase velocity $c$) and the vorticity eigenmode structure $\zeta(y)$ evolve as viscosity approaches zero. This transformation, analogous to quasi-mode behavior in \blue{homogeneous} stellar systems, requires solving the eigenvalue problem at extremely high Reynolds numbers $\rm Re$, necessitating high-precision numerical calculations.

\begin{figure}[t]
\centering
\includegraphics[width=86mm]{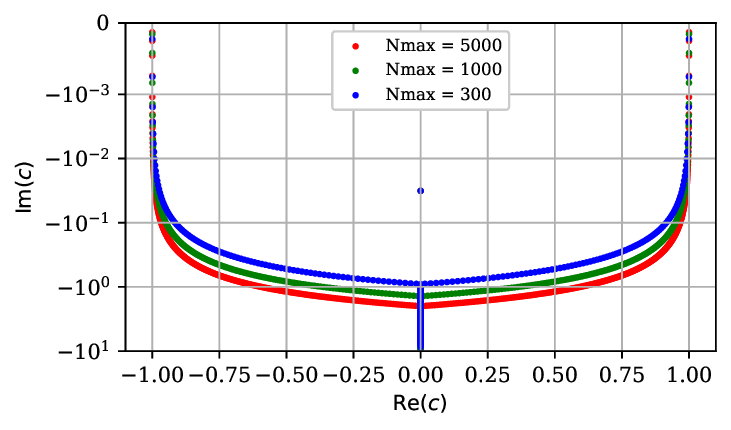}
\caption{\footnotesize Stable spectrum of eigenmodes of nearly inviscid free mixing layer $U=\tanh y$ for $k=1.05$, $\mu\equiv (k\,{\rm Re})^{-1} =10^{-4}$, obtained numerically for $N_{\rm max}=300, 1000$ and 5000. The single blue point above the triple point represents the Landau mode, which remains essentially unchanged across all $N_{\rm max}$ values (points overlaid). \blue{In purely inviscid flow the spectrum consists of van Kampen modes that cover the interval $-1\le c \le 1$ on the real $c$-axis (not marked in the figure), which generate the Landau quasi-mode.}}
\label{fig:c-spek}
\end{figure}

To explain our interest in this particular mode, we first describe the general  structure of the spectrum. The eigenvalue problem requires high-precision numerical calculations using sparse matrix methods with large dimensions $N_{\rm max}\times N_{\rm max}$. The spectrum exhibits a characteristic Y-shaped structure, where our mode of interest has smaller $|{\rm Im}\,c|$ than at the triple point (Fig.\,\ref{fig:c-spek}). The triple point position depends on $N_{\rm max}$ -- larger values shift it lower in the complex $c$-plane. As $N_{\rm max}\to\infty$, the Y-branches approach two vertical lines $c=\pm 1 - \i\mu\,k^2(1+p^2)$ with  parameter $p^2$ running from zero to infinity \red{and $\mu=\nu/k$, where $\nu$ is the kinematic viscosity.} These lines represent the continuous spectrum that appears in semi-infinite and infinite flows, where the homogeneous boundary conditions are replaced by boundedness conditions: $\psi$ and $\d\psi/dy$ bounded. The origin and expressions for the continuous spectrum in boundary layers are presented in Schmid and Henningson's textbook \cite{SH2012}. In Appendix B, we apply their derivation method to the free hyperbolic tangent flow.

As shown in Fig.\,\ref{fig:c-spek}, the viscous mixing layer spectrum includes several modes with $c_r\approx \pm 1$ and small $c_i$ values of order ${\cal O}(\mu)$. 
This means the Landau mode is no longer the least damped when viscosity is included, seemingly losing its special significance compared to plasma or stellar media. However, this conclusion is misleading. These seemingly \blue{less} stable modes dominate only at very late times during initial disturbance evolution; prior to this late-time regime, the Landau mode controls the dynamics. As in inviscid flows, power-law decay becomes prominent only after the exponentially decaying Landau disturbance has diminished by several orders of magnitude \cite{PS_22}. We verified this by examining the evolution of initial vorticity disturbances with arbitrary shapes, confirming they rapidly acquire the eigenfunction (EF) form before decaying more slowly while gradually deviating from the EF structure (see Sec.\,\ref{sec:evo} for details).

The structure of the paper is as follows. In Sec.\,\ref{sec:analytic} we analytically derive the low-viscosity eigenvalue correction to the Landau quasi-mode via perturbation theory (Sec.\,\ref{subsec:2a}) and construct an approximate analytical expression for the EF using matched asymptotic expansions (Sec.\,\ref{subsec:2b}). Sec.\,\ref{sec:num} describes the numerical procedures used to validate these analytical solutions. Specifically, Sec.\,\ref{subsec:3a} presents the derivation of the dispersion equation in matrix form, and Sec.\,\ref{subsec:3b} discusses the boundary conditions. The non-trivial numerical methods required to solve the eigenvalue problem for vanishingly small viscosity and the results are presented in Sec.\,\ref{sec:val}. The evolution of initial disturbances is examined in Sec.\,\ref{sec:evo}, and Sec.\,\ref{sec:conc} discusses our findings. Finally, Appendix\,\ref{appA} provides detailed derivations of the matrix equation with explicit forms of the matrix elements, while Appendix\, \ref{appB}  derives the continuous spectrum for the viscous free hyperbolic tagent flow.

\section{Analytical Expressions}
\label{sec:analytic}

\subsection{Mathematical Formulation and Eigenvalue Corrections}
\label{subsec:2a}
For two-dimensional disturbances of the stream function ${\hat\psi}(x,y,t)=\psi(y)\,\exp [-\i\,(\omega t-kx)]$ in inviscid incompressible plane-parallel shear flows $V_x=U(y)$ with an inflection point, only continuous van Kampen mode spectra with real frequency $\omega$ (or phase velocity $c=\omega/k$) exist in the stable wavenumber region $k>k_{\rm crit}$, with phase velocities lying in the interval $U_{\rm min}\le c\le U_{\rm max}$ (e.g., \citet{DR81, Balmforth_Morrison}). The corresponding vorticity perturbation ${\hat\zeta}(x,y,t)=\zeta(y)\,\exp [-\i\,(\omega t-kx)]$ has amplitude
\be 
\zeta(y)=\Delta\psi(y) = \bigl({\d^2}/{\d y^2}-k^2\bigr)\,\psi(y)\,. 
\label{eq:psi} 
\ee

For viscous shear flows, the Orr-Sommerfeld equation is (see, e.g., \cite{SH2012}):
\begin{equation}
(U-c)\,\Bigl(\frac{\d^2}{\d y^2}-k^2\Bigr)\,\psi-
U''\,\psi+\i\,\mu\,\Bigl(\frac{\d^2}{\d y^2}-k^2\Bigr)^2\psi=0
\label{eq:OZ}
\end{equation}
with boundary conditions (in the case of free flow) $\psi(\pm \infty)=\psi'(\pm \infty)=0$. \red{Here 
\be
\mu \equiv \nu/k = (k\,{\rm Re})^{-1}
\label{eq:mu}
\ee 
and ${\rm Re}$ is the Reynolds number.}
We consider the hyperbolic tangent free shear flow $U=\tanh y$, which has been extensively studied by many authors (see review paper by Michalke \cite{Michalke_1972} and the works cited therein), focusing on the weakly damped disturbances near the stability boundary in the stable region $0< k-1\ll 1$.

\blue{Note that plane-parallel flow $U(y)$ is not stationary in the presence of uniform viscosity. To maintain stationarity, a compensating body force $F=-\nu\,\d^2 U(y)/\d y^2$ is introduced into the governing equations (an approach known as the Jeans swindle in stellar dynamics). This forcing term does not appear in the linearized equation (\ref{eq:OZ}).}

Writing equation (\ref{eq:OZ}) in integral form:
\begin{equation}
(U\!-\!c)\,\zeta(y)-U''\int_{-\infty}^\infty\!\!\!\!\!\d y'\, G_k(y,y')\,\zeta(y')\!=\!-\i\,\mu\,\Delta\,\zeta(y),
\label{eq:OZ_int}
\end{equation}
where 
\be
G_k(y,y')=-\frac{1}{2k}\,\exp (-k|y-y'|)
\label{eq:Green_function} 
\ee
is the Green function and $\Delta\zeta \equiv \bigl({\d^2}/{\d y^2}-k^2\bigr)\,\zeta$.
\red{We treat the RHS of (\ref{eq:OZ_int}) as a small perturbation to the inviscid equation} and seek the viscosity correction ${\cal O}(\mu)$ to the Landau quasi-mode `eigenvalues' $c=c_{\rm L}(k)$ by setting
\begin{equation}
\zeta(y)=\zeta_{\rm L}(y)+\zeta_1(y), \quad c=c_{\rm L}(k)+c_1,
\end{equation}
\red{where $\zeta_1$ and $c_1$ are ${\cal O}(\mu)$.} Here $\zeta_{\rm L}(y)$ is the inviscid EF evaluated on a contour in the lower complex $y$-plane that passes below the resonance point $y=y_c$ where $U(y_c)=c_{\rm L}$ and ${\rm Im}(c_{\rm L})<0$ (see,  e.g., \cite{PS_22}). 
\red{This contour, which passes below the singular point, follows from Lin's bypass rule \cite{Lin_55}, justified by the inner problem (see Sec.\,\ref{subsec:2b}).}

Linearizing equation (\ref{eq:OZ_int}) \red{in the complex $y$-plane (the extension of the Green function to the complex plane is described in \cite{PS_22}):}
\begin{multline}
     (U-c_{\rm L})\,\zeta_1(y)-c_1\,\zeta_{\rm L}(y)\\-U''(y)\int\limits_{\curvearrowbotright}\d y' \, G_k(y,y')\,\zeta_1(y') = -\i\mu\,\Delta\zeta_{\rm L}\,,   
\end{multline}
where `${\curvearrowbotright}$' indicates integration along the same contour described above. \red{The inviscid EF $\zeta_{\rm L}(y)$ corresponding to the Landau quasi-mode is not defined on the real $y$-axis, but only on this complex contour.} We multiply both sides by the adjoint function \red{of the inviscid problem, $\zeta_{\rm L}^\dag=\zeta_{\rm L}/U''$}, which satisfies
\red{\begin{equation}
    (U-c)\,\zeta_{\rm L}^\dag(y)-\int\limits_{-\infty}^\infty \!\!\d y' \, G_k(y,y')\,U''(y')\,\zeta_{\rm L}^\dag(y') = 0,
\label{eq:adjoint} 
\end{equation}
}
and integrate on the contour:
\begin{multline}
        \int\limits_{\curvearrowbotright} \d y\, (U-c_{\rm L})\,\zeta_1(y)\,\zeta_{\rm L}^\dag\\ - \int\limits_{\curvearrowbotright}  \d y\, U''(y)\,\zeta_{\rm L}^\dag(y) \int\limits_{\curvearrowbotright}  \d y'\,G_k(y,y')\,\zeta_1(y')\\ -c_1\,\int\limits_{\curvearrowbotright}\d y\, \zeta_{\rm L}(y)\,\zeta_{\rm L}^\dag(y)=-\i\mu\,\int\limits_{\curvearrowbotright}\d y\,\zeta_{\rm L}^\dag \Delta\zeta_{\rm L}\,.
\end{multline}

Relabeling $y \leftrightarrows y'$ in the second term and using $G_k(y,y')=G_k(y',y)$:
\begin{multline}
    \int\limits_{\curvearrowbotright}\,\d y\,\zeta_1(y)\Bigl[ (U\!-\!c_{\rm L})\,\,\zeta_{\rm L}^\dag\!-\! \int\limits_{\curvearrowbotright}\d y'\, U''(y')\,\zeta_{\rm L}^\dag(y')\, G_k(y,y')\Bigr]\nonumber\\ -c_1\,\int\limits_{\curvearrowbotright}\d y\, \zeta_{\rm L}(y)\,\zeta_{\rm L}^\dag(y)=-\i\mu\,\int\limits_{\curvearrowbotright}\d y\,\zeta_{\rm L}^\dag \Delta\zeta_{\rm L}.
\end{multline}

According to the adjoint function relation (\ref{eq:adjoint}), the expression in square brackets vanishes, yielding
\be
{c_1}/{\mu}=\i\,{I_2}/{I_1}
\label{eq:c1_mu}
\ee
or, since for aperiodic modes \red{$c_1=-\i\, [\sigma(k,\mu)-\sigma_{\rm L}(k)]\equiv -\i\, \Delta\sigma(k,\mu)$},
\be {\Delta\sigma(k,\mu)}/{\mu}=-{I_2}/{I_1}.
\label{eq:Delta_sigma}
\ee
Here
\begin{equation}
I_1=\int\limits_{\curvearrowbotright} \frac{\d y}{U''(y)}\,\zeta_{\rm L}^2(y),\ \ \ \
I_2=\int\limits_{\curvearrowbotright}\frac{\d y}{U''} \,\zeta_{\rm L}\,\Delta\zeta_{\rm L}
\end{equation}

Alternative expressions for the integrals $I_1$ and $I_2$ are
\begin{equation}
I_1=\int\limits_{\curvearrowbotright} \frac{\d y}{U-c_{\rm L}}\,\psi_{\rm L}(y)\,\zeta_{\rm L}(y),\ \ \ \
\label{eq:I1}
\end{equation}
\be
I_2\!=\!\int\limits_{\curvearrowbotright}\frac{\d y}{U-c_{\rm L}}\,\psi_{\rm L}\,\Delta\zeta_{\rm L}\!=\!-k^2\,I_1\!+\!\int\limits_{\curvearrowbotright}\frac{\d y}{U-c_{\rm L}}\,\psi_{\rm L}\zeta_{\rm L}'',
\label{eq:I2}
\ee
where we used the inviscid equation relation
\begin{equation}
(U-c_{\rm L})\,\zeta_{\rm L}-U''\,\psi_{\rm L}=0.
\label{eq:inviscid}
\end{equation}

\red{Expressions (\ref{eq:I1}) and (\ref{eq:I2}) apply in both the stable and unstable regions in $k$. Note, however, that in the unstable region, $k<1$, the contour need not be deformed into the lower half-plane of $y$ and the integration can be performed along the real axis. In the marginal case, $k=1$, where $c_{\rm L}=0$ and the integrands in $I_1$ and $I_2$ contain a singularity $1/y$, integration along a complex contour can also be avoided by taking half-residue.}

As a check of expression (\ref{eq:c1_mu}), we calculate the viscosity correction to the frequency $c_{\rm L}=0$ corresponding to $k=1$ (the stability boundary). For this case,
\begin{equation}
\psi_{\rm L}=\frac{1}{\cosh y}, \ \ \zeta_{\rm L}=-\frac{2}{\cosh^3 y}.
\end{equation}
The result is known (see, e.g., \citet{Huerre_1980,Chu_Shu1987}): $c_1=-4\i\,\mu$.
From (\ref{eq:I1}, \ref{eq:I2}), we obtain $I_1=-2\pi\i$, $I_2=8\pi\i$, so Eq.~(\ref{eq:c1_mu}) yields the correct known result. Note that the precise corrected eigenvalues provided by relation (\ref{eq:Delta_sigma}) are crucial for the numerical matrix eigenvalue calculations presented in Sec.\,\ref{sec:val}, where these serve as initial approximations for the iterative solution.

We can obtain a more extended test by considering perturbations from both viscosity and deviation of $k$ from 1. This allows us to derive an expression for the phase velocity correction to $c_{\rm L}=0$ due to both viscosity and wavenumber deviation from 1.
Setting
\be
c=c_{\rm L}+c_1=c_1,\ \ \ k=k_0+k_1=1+k_1,
\ee
\be \psi(y)=\psi_{\rm L}(y)+\psi_1(y),\ \ \zeta(y)=\zeta_{\rm L}(y)+\zeta_1(y).
\ee
We have
\begin{equation}
(U-c_{\rm L})\,\zeta_1(y)-c_1\,\zeta_{\rm L}(y)-U''(y)\,\psi_1(y)=-\i\mu\,\Delta\zeta_{\rm L}.
\label{eq:zagot}
\end{equation}
We need to express $\psi_1(y)$ in integral form. From relation (\ref{eq:psi}), we obtain
\red{
\begin{equation}
\psi_1''-k_0^2\,\psi_1=\zeta_1+2 k_0 k_1\psi_{\rm L},
\end{equation}
}
and
\begin{multline}
    \psi_1(y)=\int\limits_{-\infty}^\infty\d y'\, G_{k_0}(y,y')\,\zeta_1(y')\\
    +2 k_0 k_1 \int\limits_{-\infty}^\infty \d y'\, G_{k_0}(y,y')\,\psi_{\rm L}(y')\,.
\end{multline}
Substituting $\psi_1(y)$ into (\ref{eq:zagot}), multiplying by $\zeta_{\rm L}^\dag(y)$, and integrating over $y$, after relabeling $y\leftrightarrows y'$ in double integrals containing $G_{k_0}(y,y')$, the $\zeta_1(y)$ terms cancel, yielding
\begin{multline}
    -c_1\int\limits_{\curvearrowbotright}\d y\, \zeta_{\rm L}(y)\,\zeta_{\rm L}^\dag(y)\\-2k_0 k_1\int \d y\, \psi_{\rm L}(y) \int\limits_{\curvearrowbotright} \d y'\, U''(y')\zeta_{\rm L}^\dag(y')\,G_{k_0}(y',y)\\=-\i\mu\,\int\limits_{\curvearrowbotright} \d y\,\zeta_{\rm L}^\dag \Delta\zeta_{\rm L},
\end{multline}
or, using (\ref{eq:adjoint}),
\begin{multline}
    -c_1\!\!\int\limits_{\curvearrowbotright} \d y\, \zeta_{\rm L}(y)\,\zeta_{\rm L}^\dag(y) - 2\, k_0 k_1\int\limits_{\curvearrowbotright}\! \d y\,  (U\!-\!c_{\rm L})\,\zeta_{\rm L}^\dag(y)\,\psi_{\rm L}(y)\\=-\i\mu\,\int\limits_{\curvearrowbotright}\d y\,\zeta_{\rm L}^\dag \Delta\zeta_{\rm L}.
\end{multline}
Transforming the $k_1$ term:
\begin{multline}
    \int \d y\,(U-c_{\rm L})\,\zeta_{\rm L}^\dag(y)\,\psi_{\rm L}(y)\\ =\int \d y\,\left[(U-c_{\rm L})\,\frac{\zeta_{\rm L}(y)}{U''}\right]\,\psi_{\rm L}(y)=\int \d y\, \psi_{\rm L}^2\,,
\end{multline}
where the last equality uses (\ref{eq:inviscid}). We obtain
\begin{equation}
I_1\,c_1=\i\,\mu\,I_2 -2\,k_1\,I_3,
\end{equation}
where 
$I_3=\int \d y\,\psi_{\rm L}^2=\int{\d y}/{\cosh^2 y}=2$, so
\begin{equation}
c_1=\i\,\mu\,\frac{I_2}{I_1}-2k_1\,\frac{I_3}{I_1} = -4\i\,\mu-\frac{2\i}{\pi}\,k_1.
\end{equation}
Thus, $\sigma\equiv -c_i\equiv-{\rm Im}(c)>0$  
 is
\be
\sigma=\frac{2}{\pi}\,(k-1)+4\,\mu,
\label{eq:c_vis}
\ee
yielding the boundary shift from its inviscid value: $k_{\rm crit}=1-2\pi\,\mu$. This result is also known \citep{Huerre_1980, Chu_Shu1987}. The reduction of growth rates and stability boundary shift due to viscosity in the hyperbolic tangent free mixing layer was first numerically considered by Betchov and Szewszyk \cite{Betchov_Szewszyk_1963}. Here we establish that result (\ref{eq:c_vis}) also applies to the stable region, where viscosity increases damping rates.

\subsection{Approximate Analytical Expression for Eigenfunctions}
\label{subsec:2b}

Here it is convenient to work with Eq.~(\ref{eq:OZ}) for the stream function $\psi$. We introduce the small parameter (see, e.g.\cite{Su_Oberman1968, Benney&Maslowe1975, Auerbach_77, Huerre_1980, Chu_Shu1987,ChurShukh1996, PS_2025,PS_25PRE})
\be
\delta \equiv \mu^{1/3},
\ee
the scale of the viscous critical layer, and adopt the scaling 
\be
Y \equiv y/\delta, \qquad {\mathfrak S} 
= {\sigma}/{\delta}.
\label{eq:scaling}
\end{equation}
We employ the method of matched asymptotic expansions (see, e.g. \cite{Bar_Zel_71, Malley2014}) for solving singularly perturbed differential equations. This divides the $y$-axis into two regions: the inner region where $|y|\lesssim \ell$ with $\delta\ll \ell\ll 1$, and the outer region where $|y| \gtrsim \ell$. 
Note that for finding the approximate expression for the EF, it does not matter whether we use the inviscid value $\sigma_{\rm L}$ or $\sigma(\mu)$ in scaling (\ref{eq:scaling}).

We solve the inner and outer regions separately, then match them at the overlap region $|y|\sim \ell$. This matching determines the integration constants and yields the dispersion equation (already known from the solvability condition using Lin's bypass rule in the previous subsection). Specifically, the inner asymptotics of the outer solution as $|y|\ll 1$ must match the outer asymptotics of the inner solution as $|Y|\to \infty$. 

Regarding the deviation of $k$ from the inviscid stability boundary $k=1$, we first consider the case where the deviation is small of order ${\cal O}(\delta)$:
\[
{(k-1)}/{\delta}={\bar k}_1 ={\cal O}(1),
\]
and later generalize to arbitrary (but still sufficiently small) deviations that preserve small damping rates.

\subsubsection{The case of $k-1={\cal O}(\delta)$}

From (\ref{eq:OZ}) in the inner region, we obtain
\begin{multline}
    \psi^{IV}-\i\,(Y+\i\,{\mathfrak S})\,\psi'' \\+\delta^2\,\Bigl[\i\,(Y\!+\!\i{\mathfrak S})\,\psi\!-\!2\psi''\!+\!{\cas{1}{3}}\,\i\,Y^3\,\psi''\!-\!2\,\i Y\,\psi\Bigr]\!+\!{\cal O}(\delta^3)=0.
\end{multline}
Here, primes denote derivatives with respect to $Y$. Setting 
\begin{equation}
\psi(Y)=\Psi_0(Y)+\delta^2\,\Psi_2(Y)+{\cal O}(\delta^3),
\label{eq:psi_exp}
\end{equation}
we have
\be
\Psi_0^{IV}(Y)-\i\,(Y+\i{\mathfrak S})\,\Psi_0''(Y)=0.
\ee
From matching with the outer solution (see below), we obtain 
\be
\Psi_0(Y)=P={\rm const}.
\ee
At order ${\cal O}(\delta^2)$
\begin{equation}
\Psi_2^{IV}(Y)-\i\,(Y+\i\,{\mathfrak S})\,\Psi_2''(Y)=\i\,(Y-\i\,{\mathfrak S})\,P.
\end{equation}
Setting
\begin{equation}
\Psi_2(Y)=\Bigl[-\cas{1}{2}\, Y^2+\Phi(Y)\Bigr]\,P,
\label{eq:Psi2}
\end{equation}
we obtain
\begin{equation}
\Phi^{IV}-\i\,(Y+\i\,{\mathfrak S})\,\Phi''=2\,{\mathfrak S}.
\label{eq:for_Phi}
\end{equation}

Equation (\ref{eq:for_Phi}) coincides (up to notation) with Eq.~(26) of \cite{PS_2025}:
\be
Z''-\i\,(Y+\i\,{\mathfrak S})\,Z=2\,{\mathfrak S},
\label{eq:Z}
\ee
with the solution
\begin{equation}
 \Phi''(Y)=-2\,{\mathfrak S}\,\int\limits_0^\infty \d q\, \e^{-q^3/3-\i\,q\,(Y+\i\,{\mathfrak S})}\,.
 \label{eq:Phi_SS}
\end{equation}
The solution of (\ref{eq:Z}) is obtained using the Fourier transform:
\begin{equation}
{\bar Z}(q)=\frac{1}{2\pi}\int\limits_{-\infty}^{\infty} \d Y Z(Y)\,\e^{-\i qY}\,,
\end{equation}
where the equation for ${\bar Z}(q)$ is
\begin{equation}
\frac{\d{\bar Z}}{\d q}+({\mathfrak S}-q^2)\,{\bar Z}=2\,{\mathfrak S}\,\delta(q),
\end{equation}
and $\delta(x)$ is the Dirac delta function. Its solution is
\begin{equation}
{\bar Z}=-2\,{\mathfrak S}\,\e^{q^3/3-{\mathfrak S}\,q}\,\Theta(-q),
\end{equation}
where $\Theta(x)$ is the Heaviside step function. The inverse Fourier transform yields
\begin{equation}
 Z(Y)=-2\,{\mathfrak S}\,\int_0^\infty \d q\, \e^{-q^3/3-\i\,q\,(Y+\i\,{\mathfrak S})}\,.
 \label{eq:Z_Y}
\end{equation}
So
\begin{equation}
\Psi_2''(Y)=\bigl[-{\cas{1}{2}}\,Y^2+\Phi(Y)\bigr]''\,P=\bigl[-1+\Phi''(Y)\bigr]\,P.
\end{equation}
For the vorticity, we have
\begin{multline}
    \zeta(y)=\frac{1}{\delta^2} \Psi''(Y)-\Psi(Y)\\=\frac{1}{\delta^2} \,[\Psi_0(Y)+\delta^2 \Psi_2(Y)]''-[\Psi_0(Y)+\delta^2\,\Psi_2(Y)]\\=\Psi_2''(Y)-P=[\Phi''(Y)-2]\,P+{\cal O}(\delta^2),
\label{eq:zeta_y}
\end{multline}
or, finally, 
\begin{equation}
\zeta(Y)=-2\,P\,\Bigl[\,{\mathfrak S}\,\int\limits_0^\infty \d q\, e^{-q^3/3-\i q\,(Y+\i\,{\mathfrak S})}+1\Bigr].
\label{eq:zeta_inner}
\end{equation}

For the outer asymptotics $|Y|\to\infty$ of the inner solution $\psi(Y)=P+\delta^2\Psi_2(Y)$, we have from (\ref{eq:Psi2}) and (\ref{eq:for_Phi}), 
\red{(omitting the term $\Phi^{IV}$ in  (\ref{eq:for_Phi}), and integrating the expression for $\Phi''$ twice over $Y$):}
\begin{equation}
\Psi_2\approx \Bigl[-{\cas{1}{2}}\,Y^2+2\,\i\,{\mathfrak S}\,Y\,(\ln \delta Y+C)\Bigr]\,P, 
\label{eq:asympt}
\end{equation}
where $C$ is a constant determined by matching with the outer solution. It is known (see, e.g., \cite{Huerre_1980, Chu_Shu1987}) that the asymptotics (\ref{eq:asympt}),  that follows from the asymptotics of equation $\Phi^{IV}- \i Y\,\Phi''=2\,{\mathfrak S}$, (which in turn is the asymptotic form of equation (\ref{eq:for_Phi}) at $|Y|\gg {\mathfrak S}$), is valid only in the part of the complex plane $Y$, namely,
$-\frac{7}{6}\,\pi<{\rm arg}\,Y <\frac{1}{6}\,\pi$,
including the lower half-plane. In the rest of the complex plane, solutions grow exponentially with $|Y|$. This is precisely where Lin's bypass rule \cite{Lin_55} originates. \red{More precisely, this bypass rule is called the Landau-Lin bypass rule, as it was first established by Landau \cite{Lan_46} based on the causality principle (the disturbance is absent in the distant past), and later justified by Lin using the dissipativity principle (viscosity $\nu>0$).}

\smallskip

The outer asymptotics of the inner solution must be matched with the inner asymptotics of the outer solution to establish the connection between ${\bar k}_1$, $\mu$, and ${\mathfrak S}$ (dispersion relation) and determine the amplitude parameter $P$ and constant $C$. 

\red{Note that the outer problem for the hyperbolic tangent velocity profile coincides with the outer problem in \cite{Huerre_1980, Chu_Shu1987} when nonlinear terms are neglected. Therefore, the inner asymptotics of the outer solution can be taken directly from, for example, \cite{Chu_Shu1987}. The difference between the present case and \cite{Huerre_1980, Chu_Shu1987}, besides the purely linear treatment here, is that \cite{Huerre_1980, Chu_Shu1987} studied weakly unstable disturbances with $k<1$, whereas we consider damped disturbances with $k>1$. Therefore, the inner problem requires new calculations, unlike the outer problem.}

The outer solution is constructed separately for $y>0$ and $y<0$, and these parts must be matched through the critical layer (i.e., the inner region). In the notation of this work, it can be written as 
\be
\psi^{\rm out}(Y)=\Psi^{\rm out}_0(Y)+\delta^2\Psi^{\rm out}_2(Y)+...,
\label{eq:psi_out}
\ee 
where
\be 
\Psi^{\rm out}_0(Y)=P={\rm const},
\label{eq:psi_out0}
\ee
\begin{equation}
\Psi^{\rm out}_2=\Bigl\{-{\cas{1}{2}}\,Y^2 +\Bigl[\,\i\, {\mathfrak S}\,(2\Lambda-1)-b^{\pm}\Bigr]\,Y\Bigr\}\,P,\ \ 
\label{eq:outer_asymp}
\end{equation}
where $\Lambda=\ln |\delta\, Y|$ for $\delta\ll |y|\ll 1$, with $y$ positive or negative respectively, and the jump $b^+-b^-$ is
\begin{equation}
b^+-b^-=4\,{\bar k}_1.
\end{equation}
Here $P$ is the amplitude of the inviscid neutral mode with $k=1$, $c=0$:
\be
\psi_{\rm neut} (y)=\frac{P}{\cosh y}\,.
\ee
We consider its perturbation $\psi(y)-\psi_{\rm neut}(y)$ induced by small viscosity and deviation of $k$ from the boundary, with $k=1+{\bar k}_1\,\delta$ and $c=c_1=-\i\,{\mathfrak S}\,\delta$.

Matching (\ref{eq:outer_asymp}) with (\ref{eq:asympt}) yields $C=-\frac{1}{2}(1-\i\,\pi)$, $b^+=-b^-=2{\bar k}_1$ and the connection between ${\mathfrak S}$ and ${\bar k}_1$, i.e., dispersion equation: \be{\mathfrak S}=(2/\pi)\,{\bar k}_1\ee (note that in the accepted ordering, the ${\cal O}(\mu)$ correction to the inviscid value $\sigma_{\rm L}$ is ignored compared to the ${\cal O}(\mu^{1/3})$ correction). We have already obtained this dispersion relation in Sec.\,\ref{subsec:2a}. by postulating Lin's bypass rule \cite{Lin_55}, when using the solvability condition for the correction $\zeta_1$ associated with deviations of $k$ from 1 and $c$ from zero (see (\ref{eq:c_vis})). However, this analysis explicitly demonstrates how accounting for viscosity dictates this bypass rule in the lower half-plane.

In unscaled form, the outer asymptotics of the inner solution for $\delta\ll |y|\sim\ell\ll 1$ has the form
\be
{\rm Re}(\psi) \to P\,\bigl(1\!-\!{\cas{1}{2}}\,y^2\bigr), \ \ 
{\rm Im}(\psi)\to 2\,P\,\sigma\,y\,\bigl(\ln|y|\!-\!{\cas{1}{2}}\bigr),
\ee
\be
{\rm Re}(\zeta)\to -\frac{2\,y^2}{y^2+\sigma^2}\,P, \ \ {\rm Im}(\zeta)\to \frac{2\,\sigma\,y}{y^2+\sigma^2}\,P,
\ee
and 
\be
{\rm Re}(\zeta)\to -2\,P, \ \ {\rm Im}(\zeta)\to (2\,\sigma/{y})\,P,
\label{eq:z_asymp}
\ee when $|y|\gg\sigma.$

\subsubsection{The case of arbitrary $k-1$}
Now consider the case where $k-1$ is large enough that we do not treat it as order $\delta$, but still consider the damping rate small: $\sigma={\cal O}(\delta)$. We obtain for the inner problem
\begin{multline}
    \psi^{IV}-\i\, (Y+\i\,{\mathfrak S})\,\psi''\\+\delta^2\,\Bigl[-2 k^2\,\psi''\!+\!{\cas{1}{3}}\,\i\,Y^3\psi''\! +\!\i\,k^2\,(Y\!+\!\i\,{\mathfrak S})\,\psi -2\i Y\psi\Bigr]\\+{\cal O}(\delta^3)
\end{multline}
We again seek $\psi(Y)$ in the series form
\begin{equation}
\psi(Y)=\Psi_0(Y)+\delta^2\,\Psi_2(Y)+{\cal O}(\delta^3).
\end{equation}
We have $\Psi_0(Y)=P$ and
\begin{equation}
\Psi_2^{IV}-\i\,(Y+\i\,{\mathfrak S})\,\Psi_2''=-\i\, k^2(Y+\i\,{\mathfrak S})\,P+2\i Y\,P.
\end{equation}
Setting
\begin{equation}
\Psi_2=\left[\cas{1}{2}\,(k^2-2)\,Y^2+\Phi(Y)\right]\,P,
\end{equation}
we obtain for $\Phi(Y)$ the same equation (\ref{eq:for_Phi}):
\begin{equation}
\Phi^{IV}-\i\,(Y+\i\,{\mathfrak S})\,\Phi''=2{\mathfrak S}
\end{equation}
with the same solution (\ref{eq:Phi_SS}) for $\Phi''$. For $\Psi_2''(Y)$, we have
\begin{equation}
 \Psi_2''=\left[(k^2-2)+\Phi''\right]\,P.
 \end{equation}
and for $\zeta(y)$:
\begin{multline}
 \zeta(y)=\frac{\d^2\psi}{\d y^2}-k^2\psi=\frac{1}{\delta^2}\frac{\d^2}{\d Y^2}[\Psi_0(Y)+\delta^2 \Psi_2(Y)]\nonumber\\-k^2\,[\Psi_0(Y)+\delta^2\,\Psi_2(Y)]=\frac{\d^2\Psi_2}{\d Y^2}-k^2\,P+{\cal O}(\delta^2)\nonumber\\=\left[(k^2-2)+\Phi''(Y)\right]\,P-k^2\,P=[\Phi''(Y)-2]\,P.\phantom{--}
\end{multline}
Again we obtain the same expression (\ref{eq:zeta_y}) for $\zeta(y)$, and hence (\ref{eq:zeta_inner}).

\bigskip

\subsubsection{The structure of the resonance region at moderately large  \texorpdfstring{${\mathfrak S}$}{S}}

\red{Small viscosity plays its most important role near the resonance, where the eigenfunction structure is especially sensitive to small changes in the medium parameters. Far from resonance, small viscosity does not affect the eigenfunction structure. Recall that in the stable $k$-region, although the eigenfunction does not exist on the real $y$-axis in the purely inviscid case, the complex contour on which it does exist can be chosen to coincide almost entirely with the real axis, except near the resonance where it must pass below the singular point. A similar situation occurs in analogous plasma problems, where rare collisions produce their most significant effects near resonance (see \cite{Ng1999,PS_2025}). We therefore focus on the eigenfunction structure in the resonance region.}

\smallskip

The analytical expression (\ref{eq:zeta_inner}) for the inner-region EF $\zeta(y)$ allows us to estimate the resonant region's half-width for ${\mathfrak S} \gtrsim 1$. We find that the half-width, when expressed in the unscaled variable $y=Y\delta$, depends mainly on the eigenvalue
$\sigma$ (approximately $1.6\,\sigma\ldots 2 \sigma$, as confirmed numerically) and weakly on $\delta$. We can also estimate the oscillation period and maximal relative amplitude.

Consider the exponential modulus in the integrand, $\exp({\mathfrak S} q-\frac{1}{3}\,q^3)$, which peaks at $q=q_*\equiv\sqrt{{\mathfrak S}}$ with a value of $\exp(\frac{2}{3}\,{\mathfrak S}^{3/2})$. Expanding near the maximum, we have
\begin{equation}
\exp\bigl({\mathfrak S} q-{\cas{1}{3}}\,q^3\bigr)\approx\exp \left[\,{\cas{2}{3}}\,{\mathfrak S}^{3/2}-{\mathfrak S}^{1/2}(q-q_*)^2\right].
\end{equation}
The integration over $q$ then yields (with $p=q-q_*$)
\begin{multline}
 I=\int_0^{\infty}\d q\,\e^{-\i\,q\,Y}\,\e^{-q^3/3+{\mathfrak S} q}\approx \exp\left({\cas{2}{3}}\,{\mathfrak S}^{3/2}-\i q_*Y\right) \\
 \times \int_{-q_*}^\infty \d p \,\exp\left(-{\mathfrak S}^{1/2} p^2-\i\,p\,Y\right)\,.
\end{multline}
The lower limit, $p=-q_*=-{\mathfrak S}^{1/2}$, can be extended to $-\infty$ since the added interval's contribution is negligible for ${\mathfrak S} \gtrsim 1$. We have
\[
\int_{-\infty}^\infty \d p \,\exp (-{\mathfrak S}^{1/2} p^2- \i\,p\,Y)
=\frac{\sqrt{\pi}}{{\mathfrak S}^{1/4}}\,\exp\Bigl(-\frac{Y^2}{4\,{\mathfrak S}^{1/2}}\Bigr).
\]
Thus, finally
\be
\zeta(Y)\!=\!-2P\Bigl[\sqrt{\pi}\,{{\mathfrak S}^{3/4}}\,\e^{ \case{2}{3}\,{\mathfrak S}^{3/2}-\case{1}{4} Y^2\,{\mathfrak S}^{-1/2}-\i \sqrt{{\mathfrak S}}\,Y}\!\!+\!1\Bigr].
\label{eq:h_res}
\ee
The expression (\ref{eq:h_res}) for $\zeta(Y)$ contains a factor $\exp(\frac{2}{3}\,{\mathfrak S}^{3/2}-\frac{1}{4}\,Y^2 {\mathfrak S}^{-1/2})$ describing the envelope. This factor is large at $Y=0$ and decreases exponentially with increasing $|Y|$. We define the resonant region as the area around $Y=0$ where this factor exceeds unity:
\begin{equation}
\frac{2}{3}\,{\mathfrak S}^{3/2}-\frac{Y^2}{4\,{\mathfrak S}^{1/2}}>0.
\end{equation}
This yields a half-width of $\Delta Y=\sqrt{{8}/{3}}\,{\mathfrak S}\approx 1.63\,{\mathfrak S}$, or
\begin{equation}
    \Delta y\approx 1.63\,\sigma
    \label{eq:hw}
\end{equation} in ordinary variables $y=Y\,\delta$ and $\sigma={\mathfrak S}\,\delta$. This half-width is consistent with EF plots using a logarithmic scale for the ordinate axis. The coefficient 1.63 increases slightly when considering the prefactor, with a small dependence on $\delta$ (see Table\,\ref{tab1}).

The oscillation period $T_{y}$ in $y$ is found from the exponent $\exp (-\i \sqrt{{\mathfrak S}} Y)$:
\begin{equation}
T_y=2\pi\,\sqrt{{\mu}/{\sigma}}.
\label{eq:per}
\end{equation}
From (\ref{eq:h_res}), the resonant peak height ${\mathfrak P}$ of the EF $\zeta(y)$  is
\begin{multline}
    {\mathfrak P}(\mu)\approx
2\,\sqrt{\pi}\,{\mathfrak S}^{3/4}\,|P|\,\exp\left(\cas{2}{3}\,{\mathfrak S}^{3/2}\right)\\=2\,\sqrt{\pi}\,|P|\,(\sigma^3/\mu)^{1/4}\,
\exp\,\left(\cas{2}{3}\,\sqrt{\sigma^3/\mu}\right),
\label{eq:peak}
\end{multline}
growing without limit as $\mu \to 0$. \red{The unbounded growth of the peak height of the normalized vorticity EF indicates the disappearance of the EF in the purely inviscid case. Additional evidence for this disappearance follows from the behavior of the expansion coefficients $A_n$ in the Legendre polynomial basis $P_n$, which is used in our numerical matrix method. As viscosity decreases, the series for the EF becomes divergent (see Section~\ref{sec:val} and Figure~\ref{fig:A}).}

\section{Numerical Approach to the Eigenvalue Problem}
\label{sec:num}

\subsection{Matrix Formulation}\label{subsec:3a}

We solve the eigenvalue problem by expanding the unknown function $\psi(y)$ over a complete set of suitable functions and reducing the problem to solving an eigenvalue problem for a system of linear equations in the expansion coefficients. For free flow $U=\tanh y$, it is convenient to transform
from variable $y$ to variable $u=\tanh y$. Since the Laplace operator $\Delta=\d^2/\d y^2-k^2$ appearing in the Orr-Sommerfeld equation (\ref{eq:OZ}) becomes $\Delta=(1-u^2)\,{\cal L} -k^2$ in the new variables, where
\be
   {\cal L}=\frac{\d}{\d u}\,(1-u^2)\,\frac{\d}{\d u},
\ee
and since this operator ${\cal L}$ appears in the equation for Legendre polynomials $P_n(u)$:
\be
{\cal L}\,P_n(u)=-n(n+1)\,P_n(u),
\label{eq:Legendre}
\ee
it is natural to choose Legendre polynomials $P_n(u)$ as basis functions for expanding $\psi(u)$. This set is complete and orthogonal:
$\int_{-1}^1 \d u\, P_n(u)\,P_m(u)= 2/(2n+1)\,\delta_{mn}$.
\red{This parallels Ng {\it et al.}  \cite{Ng1999} and our earlier plasma and homogeneous stellar studies \cite{PS_2025,PS_25PRE}, where a Maxwellian equilibrium distribution naturally selects Hermite polynomials as the velocity-space basis. In general, the unperturbed profile determines the optimal basis functions and ensures sparse matrix structure.}
Equation (\ref{eq:OZ}) in variable $u$ is
\begin{multline}
c\,\left[(1-u^2)\,{\cal L}\psi-k^2\psi\right]\\= u\,\left[(1-u^2)\,{\cal L}\psi-k^2\psi\right]+2u\,(1-u^2)\,\psi\\+
\i\mu\,\left[(1-u^2)\,{\cal L}-k^2\right]\left[(1-u^2)\,{\cal L}\psi-k^2\psi\right].
\label{eq:OS_u}    
\end{multline}
Substitution
\be
\psi=\sum\limits_{n=0}^\infty A_n\,P_n(u)
\ee
yields
\begin{multline}
    c\,\sum\limits_n A_n\,\left[-n(n+1)(1-u^2)-k^2\right]\,P_n\\
=-\sum\limits_n A_n\,\left[(n-1)(n+2)\,u\,(1-u^2)+k^2 u\right]P_n \\
-\i\mu \Bigl[(1\!-\!u^2)\,{\cal L}\!-\!k^2\Bigr]\!\sum\limits_n\! A_n\!\left[n(n\!+\!1)(1\!-\!u^2)\!+\!k^2\right]\!P_n.
\label{eq:OZ_P_n}
\end{multline}
Using relations for the constructions $u\,P_n(u)$, $u^2\,P_n(u)$, and $u^3\,P_n(u)$ (see, e.g., \cite{Gradshteyn_8}), as well as equation (\ref{eq:Legendre}) for Legendre polynomials and equating coefficients of the same $P_n$, we obtain the linear homogeneous system for $A_n$:
\be
 c\,\sum_m {\cal R}_{nm}\,A_m =\sum_m{\cal Q}_{nm}\,\,A_m+\i\mu\,\sum_m{\cal T}_{nm}\,\,A_m,
 \label{eq: matrix1}
\ee
where the explicit forms of matrices ${\cal R}_{nm}$, ${\cal Q}_{nm}$, and ${\cal T}_{nm}$ are given in the Appendix A. Note that ${\cal R}_{nm}$ has 3 non-zero diagonals ($m=n, n\pm 2$), ${\cal Q}_{nm}$ has 4 non-zero diagonals ($m=n\pm 1, n\pm 3$), and ${\cal T}_{nm}$ has 5 non-zero diagonals ($m=n, n\pm 2, n\pm 4$). These sparse matrices enable high accuracy in calculating eigenvalues and EFs for vanishingly small values of $\mu$.

For aperiodic modes, ${\rm Re}(c)=0$, the system reduces to a purely real form. Making the substitution
\be
c=-\i\sigma, \ \ A_n=\i^n\, {\bar A}_n\,.
\ee
Eq.~(\ref{eq: matrix1}) becomes
\be
\sigma \sum_m{\bar{\cal R}}_{nm}\,{\bar A}_m=\sum_m {\bar{\cal Q}}_{nm}\,{\bar A}_m+\mu \sum_m{\bar{\cal T}}_{nm}\,{\bar A}_m\,,
\label{eq:matrix_bar}
\ee
where the barred matrices are also sparse with the same non-zero elements as the original matrices:
\be
{\bar{\cal R}}_{n,n}=-{\cal R}_{n,n},\ \ \ {\bar{\cal R}}_{n,n\mp 2}={\cal R}_{n,\mp 2},
\ee
\be
{\bar{\cal Q}}_{n,n\mp 1}=\mp {\cal Q}_{n,n\mp 1},\ \ {\bar{\cal Q}}_{n,n\mp 3}=\pm {\cal Q}_{n,n\mp 3},
\ee
\be
{\bar{\cal T}}_{n,n}\!=\!{\cal T}_{n,n},\  {\bar{\cal T}}_{n,n\mp 2}\!=\!-{\cal T}_{n,n\mp 2}, \  {\bar{\cal T}}_{n,n\mp 4}\!=\!{\cal T}_{n,n\mp 4}\,.
\ee

Once we know the coefficients ${\bar A}_n$ and stream function 
\be
  \psi(u)=\sum\limits_{n=0}^\infty \i^n {\bar A}_n\,P_n(u),
  \label{eq:A2psi}
\ee
we can calculate the vorticity:
\begin{multline}
    \zeta(u)=(1-u^2)\,{\cal L}\,\psi-k^2\psi= 
    \\-(1-u^2)\sum\limits_{n=0}^\infty \i^{n}\,n\,(n+1)\,{\bar A}_n P_n(u)\!-\!k^2\psi(u).
    \label{eq:A2zeta}
\end{multline}
By analogy with \cite{PS_22}, we normalize the EF by requiring the total vorticity across the flow to be equal to unity:
$
\int_{-\infty}^\infty \zeta(y)\,\d y=1
$.

\subsection{Boundary Conditions}\label{subsec:3b}
The Orr-Sommerfeld equation requires four boundary conditions: $\psi(\pm\infty)=\psi'(\pm\infty)=0$. At first glance, it seems we must add four additional equations to system (\ref{eq:matrix_bar}) to satisfy these conditions. Indeed, each basis function $P_n(u)$ does not vanish at $u=\pm 1$ (corresponding to $y=\pm\infty$): $P_n(1)=1$, $P_n(-1)=(-1)^n$. This suggests that at least two equations must be added to satisfy the first pair of boundary conditions, $\psi(u=\pm 1)=0$:
\be
  \sum_{m=0}^{{\rm N}_{\rm max}} (-1)^m {\bar A}_{2m}=0,\quad \sum_{m=0}^{{\rm N}_{\rm max}} (-1)^m {\bar A}_{2m+1}=0.
  \label{eq:bar_A}
\ee

However, it can be shown that all boundary conditions are satisfied automatically in our numerical procedure, where $\psi$ is represented as an expansion in $P_n(u)$. 

\smallskip

\blue{Setting $u=\pm 1$ in Eq.~(\ref{eq:OS_u}) and removing all terms containing the factor $(1-u^2)$, we obtain
\be
-(c\mp1)\,k^2\,\psi(\pm 1)=\i\,\mu\,k^4\,\psi(\pm 1).
\ee
This means that if $c\ne \pm 1-\i\mu\,k^2$, then $\psi(\pm 1)=0$.}

\smallskip

\blue{The second pair of boundary conditions, $\d\psi/\d y=0$ at $y=\pm\infty$, is satisfied automatically. This follows from the relation $\d\psi/\d y=(1-u^2)\,\d\psi/\d u$, which gives $[\d\psi/\d y]_{\pm\infty}=0$ when $u=\pm 1$. (Note that the derivative $[\d\psi/\d u]_{u=\pm 1}$ is always bounded due to the finite truncation $N_{\rm max}$ in the numerical procedure, so multiplying it by the factor $(1-u^2)$, which vanishes exactly, yields zero).}

\medskip

\blue{Thus, no additional equations are needed when using the $P_n(u)$ expansion.} Our numerical calculations confirm  the vanishing of $\psi$ and $\d\psi/\d y$ at $y=\pm\infty$ with very good accuracy for sufficiently large $N_{\rm max}$.

\section{Numerical Validation of the Analytical Results}
\label{sec:val}

\begin{figure} [ht]
\centering
\includegraphics[width=87mm, trim={7pt 9pt 0pt 0pt}, clip]{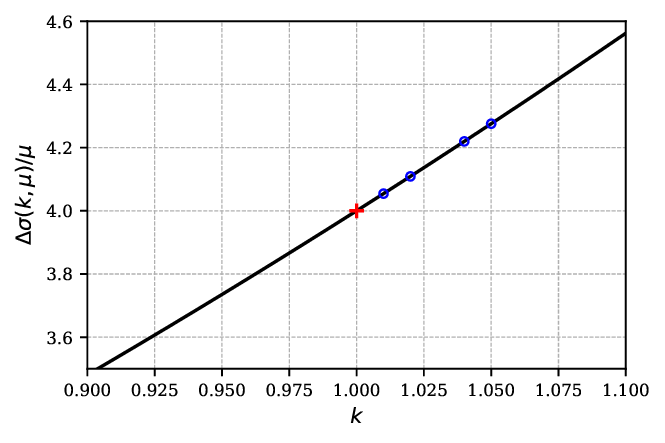}
\caption{\footnotesize The ratio  $\Delta\sigma(k,\mu)/\mu\equiv [\sigma(k,\mu)-\sigma_{\rm L}(k)]/\mu$,  where $\sigma_{\rm L}(k)\equiv \gamma_{\rm L}(k)/k$ and $\gamma_{\rm L}(k)$ is Landau damping rate,  as a function of wavenumber $k$, computed using Eq.(\ref{eq:Delta_sigma}) with contour deformation into the lower half-plane. The red plus marks the analytical result at $k=1$; blue circles show inviscid limit values given in Table \ref{tab1}.}
\label{fig:S}
\end{figure}

\begin{figure}[t]
\centering
\includegraphics[width=87mm, trim={7pt 9pt 0pt 0pt}, clip]{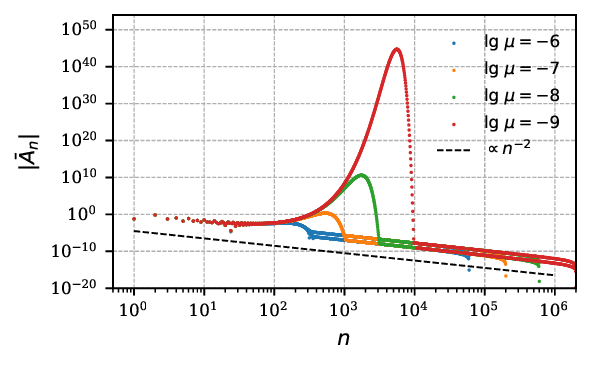}
\caption{\footnotesize Modulus of expansion coefficients $|\bar A_n|$, $n \geq 1$, for the Landau quasi-mode with $k=1.05$ at four $\mu$ values. $\bar A_0 = 1$ for all $\mu$. $|\bar A_n|$ peaks at $n_* \approx (\sigma/\mu)^{1/2}$ for very low $\mu$.
}
\label{fig:A}
\end{figure}

\begin{figure*}[t]
\centering
\includegraphics[width=150mm, trim={7pt 9pt 0pt 0pt}, clip]{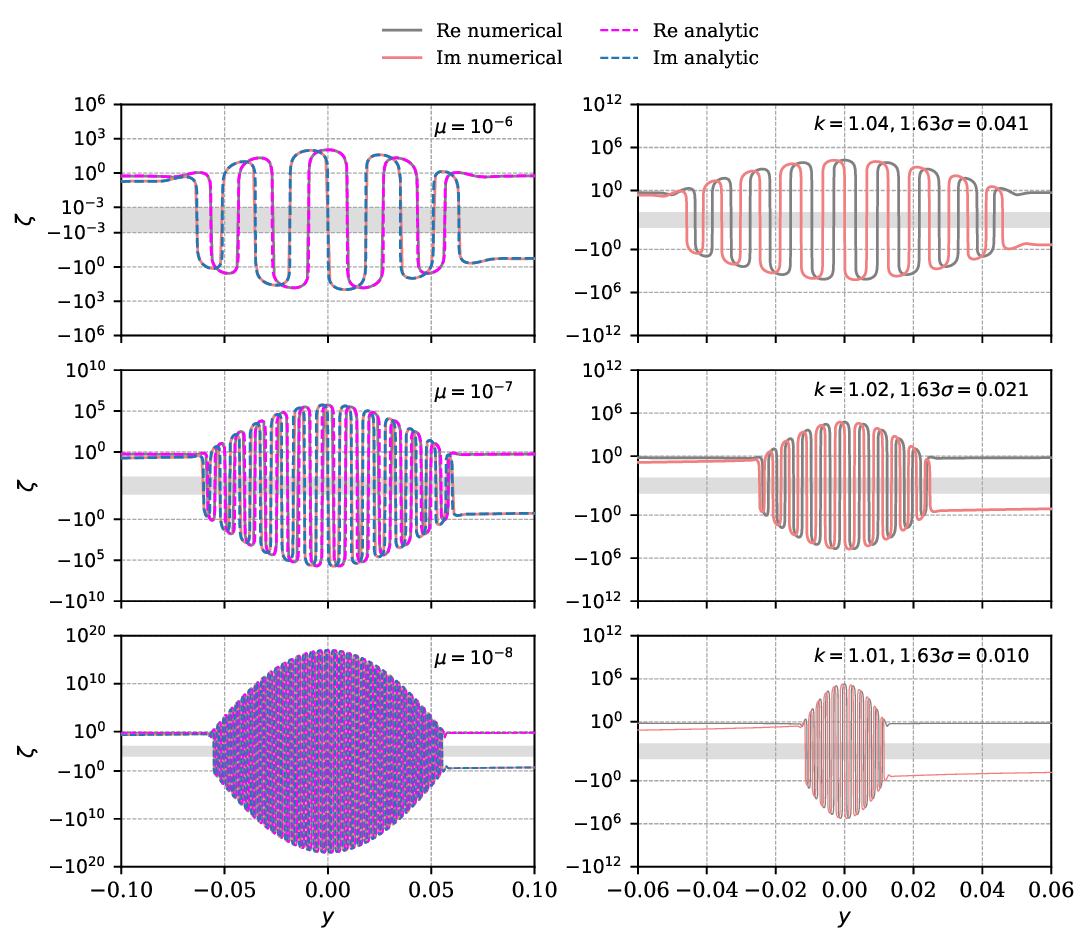}
\caption{\footnotesize Least-damped mode EFs for varying $\mu$ and $k$ values, comparing numerical solutions (solid) with analytical approximations (dashed) and nonresonant solutions (dotted). Left: EF deformation for $k=1.05$ as $\mu$ decreases (solid: numerical solutions from (\ref{eq:A2zeta}), dashed: analytical approximation (\ref{eq:zeta_inner})). 
Right: Resonance range width dependence on $\sigma$ for three $k$ values, supporting our analytical result (\ref{eq:hw}). EFs are plotted using symlog scaling (logarithmic with linear scaling near zero, indicated by gray shading).
}
\label{fig:E}
\end{figure*}

Our numerical validation of the analytical results is presented in Tab.\,\ref{tab1} and Figs.\,\ref{fig:S}--\ref{fig:E}.


Fig.,\ref{fig:S} shows the viscous correction to the Landau damping rate in the small viscosity limit. The calculation involves solving the eigenvalue problem for the function $\xi_{\rm L}(y) \equiv \Delta\zeta_{\rm L}$ derived from (\ref{eq:OZ_int}) on a complex $y$-contour deformed into the lower half-plane below the target eigenvalue. After finding $\xi_{\rm L}(y)$ and $\zeta_{\rm L}(y)$, integrals (\ref{eq:I1}, \ref{eq:I2}) can be calculated, and then the correction is calculated using (\ref{eq:Delta_sigma}). 
We observe that for $k=1$, $\Delta\sigma/\mu = 4$ (red plus), consistent with the known analytical result from \cite{Huerre_1980, Chu_Shu1987}. Blue circles show the values for models in Table~\ref{tab1}.

To obtain the eigenvalue for finite viscosity, we need to solve (\ref{eq:matrix_bar}) numerically using the precise eigenvalue correction from Sec.\,\ref{subsec:2a}, Eq.~(\ref{eq:Delta_sigma}), as the initial approximation. This system must first be truncated by setting ${\bar A}_n=0$ for $n>N_{\rm max}$. Following the previous section, we did not apply any special boundary conditions. 

\red{To obtain the eigenvalue for finite viscosity, we solve the truncated generalized eigenproblem (Eq.~(\ref{eq:matrix_bar})) using the first-order viscous correction (Eq.~(\ref{eq:Delta_sigma})) to set a real target value $\sigma$. We truncate by setting $\bar A_n=0$ for $n>N_{\rm max}$ and compute a single eigenvalue/eigenvector near $\sigma$ with MATLAB’s \texttt{eigs}, which implements an implicitly restarted Arnoldi method \cite{lehoucq1998arpack,golub2013matrix}. In our setting, $N_{\rm max}=2\times 10^5$ in 128-bit arithmetic (Advanpix) was sufficient to resolve the target eigenpair for $\mu \gtrsim 10^{-7}$.}

\red{For smaller $\mu$, we implemented the eigenvalue search in C using the GNU Multiple Precision Floating-Point Reliable Library. We used a standard shift-and-invert inverse iteration: form $\bar{\cal S}=\bar{\cal Q}+\mu\,\bar{\cal T}$, factor $\bar{\cal S}-\sigma\,\bar{\cal R}$ with a banded LU, solve $(\bar{\cal S}-\sigma\,\bar{\cal R})\,w=\bar{\cal R}\,v$ with normalization, and update the eigenvalue by the Rayleigh quotient until convergence \cite{golub2013matrix,saad2011eigs}. This targets eigenvalues near the real shift $\sigma$ and allowed $N_{\rm max}=2\times 10^6$ with 1024-bit precision, reaching $\mu=10^{-9}$.}

\red{As a separate, preliminary assessment of the problem, we also solved the Orr–Sommerfeld equation (Eq.~(\ref{eq:OZ})) by Chebyshev collocation on the mapped Chebyshev–Gauss–Lobatto grid
\be
  y=\frac{1}{s}\,{\rm artanh}(y_{\rm CGL}),\qquad y_{\rm CGL}\in[-1,1],
\ee
with fixed $s=1/4$. Dense Chebyshev differentiation matrices obtained by differentiating the global interpolant at the CGL nodes deliver spectral accuracy for smooth data \cite{trefethen2000}. Using up to 6000 collocation points and multiprecision arithmetic (Advanpix), this calculation provided consistent eigenvalues for $\mu \gtrsim 10^{-7}$ and served as an independent check of the matrix approach.}

Fig.\,\ref{fig:A} illustrates the behavior of the the expansion coefficients ${\bar A}_n$ in the stable case $k=1.05$ as $\mu \to 0$, approaching the inviscid limit. All components are real, with alternating signs from $n=1$. The absolute values $|{\bar A}_n|$ increase, reaching a maximum at $n_*\approx (\sigma_{\rm L}/\mu)^{1/2}$. This suggests a convergence issue with the series in (\ref{eq:A2psi}, \ref{eq:A2zeta}) in the inviscid limit, reflecting the absence of \red{EFs on the real $y$-axis} in the purely inviscid case. Outside the maximum region, the coefficients of $|{\bar A}_n|$ drop as $n^{-2}$, almost up to the truncation limit $N_{\rm max}$.

Figs.\,\ref{fig:E} and Table\,\ref{tab1} present our numerical validation of the analytical results. The figures compare numerical solutions for the normalized EFs $\zeta(y)$ of the Landau mode (\ref{eq:A2zeta}) (solid lines) with the analytic expression (\ref{eq:zeta_inner}) (dashed colored lines). In the non-resonant region, the solution matches with $\zeta(y)$ given by (\ref{eq:z_asymp}). This expression coincides with the inner asymptotic expansion of the outer solution, which follows from (\ref{eq:psi_out0}) and (\ref{eq:outer_asymp}).
The EFs exhibit symmetry: the real part is even, and the imaginary part is odd. We use symlog scaling for $y$-axes, combining logarithmic and linear scaling near zero, with linear regions indicated by gray shading.

Table\,\ref{tab1} presents all parameters of the resonance solutions for the models shown in Fig.\,\ref{fig:E}. Each block corresponds to a fixed wavenumber $k$. The first row gives inviscid data: Landau eigenvalue $\sigma_{\rm L}$, analytical correction $\Delta\sigma/\mu$ (\ref{eq:Delta_sigma}), and half-width $\Delta y$ (\ref{eq:hw}). Subsequent rows show eigenvalue calculations for various $\mu$ values, with half-widths $\Delta y$ determined numerically as half the distance between the rightmost and leftmost zeros of ${\rm Im}\,\zeta$ (gray cells). The analytical resonance peak values ${\mathfrak P}$ (\ref{eq:peak}) agree with numerical values within 3\%, and the analytical spatial period $T_y$ values along the $y$-coordinate (\ref{eq:per}) are completely consistent with numerical values.

The left panels in Fig.\,\ref{fig:E} show EFs $\zeta(y)$ for decreasing $\mu$ at fixed $k=1.05$. In the outer region, numerical solutions follow the nonresonant solution but begin oscillating near resonance as $|y|$ approaches zero. We observe four oscillations for $\mu=10^{-6}$ (top), 12 oscillations for $\mu=10^{-7}$ (middle), and numerous oscillations for $\mu=10^{-8}$ (bottom), illustrating the absence of EFs in the inviscid limit.

The right panels show the dependence of the width of resonance regions on damping rates for three EFs $\zeta(y)$ with damping rates $\gamma=k\,\sigma$ in the approximate ratio 0.025\,:\,0.013\,:\,0.006. The half-width of the resonance region approaches the theoretical result as $\mu$ vanishes, as confirmed by the data in Table\,\ref{tab1}.

\begin{table*}
\setlength{\tabcolsep}{6pt}
\begin{tabular}{ccccccccc}
\toprule
$k$                       & $\log_{10}\mu$ &$\sigma$                 & $\Delta\sigma/\mu$  & $\delta$  & ${\mathfrak S}$  & $\Delta y$    & ${\mathfrak P}    $ & $T_y$ \\ [2pt]
\midrule
\multirow[t]{5}{*}{1.01}  & ---      & $0.00635\;00336\;32369$ & $4.0543$            & ---        & ---      &      $0.0104$ & ---                 & ---     \\
                          & ${-9}$   & $0.00635\;00376\;86505$ & $4.0541$            & $0.001$    & $6.35$   & \cgr $0.0114$ & $1.972\cdot 10^{5}$ & 0.00249 \\ [2pt]
\midrule
\multirow[t]{5}{*}{1.02}  & ---      & $0.01266\;81081\;94502$ & $4.1089$            & ---        & ---      &      $0.0206$ & ---                 & ---     \\
                          & ${-8}$   & $0.01266\;81492\;82855$ & $4.1088$            & $0.00215$  & $5.88$   & \cgr $0.0247$ & $5.795\cdot 10^{4}$ & 0.00558 \\ [2pt]
\midrule
\multirow[t]{5}{*}{1.04}  & ---      & $0.02521\;05473\;48382$ & $4.2196$            & ---        & ---      &      $0.0411$ & ---                 & ---     \\
                          & ${-7}$   & $0.02521\;09693\;02175$ & $4.2195$            & $0.00464$  & $5.43$   & \cgr $0.0458$ & $1.863\cdot 10^{4}$ & 0.01251 \\   [2pt]              
\midrule
\multirow[t]{5}{*}{1.05}  & ---      & $0.03143\;59666\;77492$ & $4.2756$            & ---        & ---      &      $0.0512$ & ---                 & ---     \\
                          & ${-8}$   & $0.03143\;60094\;32824$ & $4.2755$            & $0.00215$  & $14.59$  & \cgr $0.0553$ & $1.160\cdot 10^{17}$ & 0.00354 \\                 
                          & ${-7}$   & $0.03143\;63942\;30497$ & $4.2755$            & $0.00464$  & $6.77$   & \cgr $0.0601$ & $6.029\cdot 10^{5}$ & 0.01121 \\                 
                          & ${-6}$   & $0.03144\;02421\;66315$ & $4.2755$            & $0.01$     & $3.14$   & \cgr $0.0632$ & $1.099\cdot 10^{2}$ & 0.03543 \\                 
\bottomrule
\end{tabular}
\caption{Characteristics of the Landau mode vorticity in the resonant region. Each block corresponds to a $k$ value. The first row gives inviscid characteristics: Landau eigenvalue $\sigma_{\rm L}(k)$, analytical viscous correction (\ref{eq:Delta_sigma}), and half-width (\ref{eq:hw}).
Subsequent rows show finite-$\mu$ results: parameters $\sigma(k,\mu)$, $\Delta\sigma(k,\mu)/\mu$,   $\delta$,  ${\mathfrak S}$, resonance half-width $\Delta y$, resonance peak ${\mathfrak P}$ (\ref{eq:peak}) and spatial oscillation period $T_y$ along the $y$-coordinate (\ref{eq:per}). Gray cells indicate numerical values.} 
\label{tab1}
\end{table*}

\section{Evolution of Initial Disturbances}
\label{sec:evo}

As noted in the Introduction, viscosity eliminates van Kampen modes but introduces many new modes in addition to the true Landau mode. Some of these decay more slowly than the Landau mode, with $c_r = \pm 1$, corresponding to oscillating modes with frequency $\omega = |c_r| k = k$ and period $T_{\rm t} = 2\pi/k$. Therefore, the Landau mode is no longer the least damped, unlike in plasma and homogeneous stellar media where it remains least damped even with rare collisions. This means that if the initial perturbation contains other modes, the final asymptotics will be determined by the more slowly decaying components rather than the Landau mode.

The question arises: over sufficiently long times after initial disturbance onset, does the vorticity decay with the Landau mode damping rate and maintain the EF form? A positive answer would justify our focus on this particular mode.

To address this question, we consider the initial value problem. Two approaches are possible. The first decomposes the initial perturbation into EFs and traces evolution by summing contributions at time $t$ -- a laborious procedure requiring calculation of adjoint functions for expansion coefficients. The second approach is simpler: directly solve the Orr-Sommerfeld equation (\ref{eq:OZ}) with the replacement $c \to (\i/k)\, \d/\d t$. We implement this using our matrix representation for expansion coefficients ${\bar A}_n(t)$, then calculate the corresponding sum to find the vorticity $\zeta(y,t)$.

Setting $\sigma = -(1/k)\, \d/\d t$ in the matrix equation (\ref{eq:matrix_bar}), we obtain a system of first-order linear ordinary differential equations:
\begin{equation}
\sum_{m=0}^{N_{\rm max}}\! {\bar{\cal R}}_{nm} \frac{\d{\bar A}_m}{\d t}\!=\! -\sum_{m=0}^{N_{\rm max}}\! {\bar{\cal Q}}_{nm} {\bar A}_m\! -\! \mu \sum_{m=0}^{N_{\rm max}}\! {\bar{\cal T}}_{nm} {\bar A}_m,
\end{equation}
where $n = 0, 1, \ldots, N_{\rm max}$, which can be solved using the Euler method. For initial conditions, we use the stream function
\begin{equation}
\psi(u, t=0) = C(K) (1-u^2)^{K+1},
\end{equation}
or equivalently $\psi(y,0) = C(K)/(\cosh y)^{2K+2}$, with corresponding initial vorticity
\begin{equation}
\zeta(u,0)\! = \!\Bigl\{2\,(K\!+\!1)\,\bigl[(2K\!+\!3)\,u^2\!-\!1\bigr] - k^2\Bigr\}\,\psi(u,0).
\label{eq:zeta_ini}
\end{equation}
Here $C(K)$ is a normalization constant determined from the condition $\int_{-\infty}^\infty \d y\, \zeta(y,0) = 1$:
\begin{equation}
C(K) = -\frac{1}{k^2} \frac{\Gamma(K+\frac{3}{2})}{\Gamma(K+1)\,\Gamma(\frac{1}{2})},
\end{equation}
where $\Gamma(x)$ is the gamma function. For integer $K$:
\begin{equation}
C(K) = -\frac{1}{k^2} \frac{(2K+1)!!}{2^{K+1} K!}.
\end{equation}

For such initial disturbances, only even coefficients are nonzero:
\begin{equation}
{\bar A}_{2m}(0) = (-1)^m \frac{4m+1}{2} \int_{-1}^1 \d u\, P_{2m}(u) \psi(u,0), 
\end{equation}
and for integer $K$, there are finitely many: ${\bar A}_0, {\bar A}_2, \ldots, {\bar A}_{2K+2}$ ($K+2$ coefficients). To illustrate the evolution of $\zeta(y,t)$, we choose an initial perturbation with $K=2$ and select $k = 1.05$ and $\mu = 10^{-4}$.

The total vorticity 
\begin{equation}
{\cal N}(t) = \int \d y \, \tilde{\zeta}(y,t),
\label{eq:calN}
\end{equation}
where $\tilde{\zeta}(y,t)$ is the amplitude at fixed $k$ defined by $\hat{\zeta}(x,y,t) = \tilde{\zeta}(y,t) \exp(\i k x)$, decays exponentially at damping rate $\gamma(k,\mu)\approx \gamma_{\rm L}(k)$: ${\cal N} \propto \exp(-\gamma\,t)$. This quantity is analogous to perturbed electron density in Landau's classical plasma problem \cite{Lan_46} (for the plasma-hydrodynamics analogy, see \cite{AF_79}).

\begin{figure} [htbp]
\centering
   \includegraphics[width=86mm]{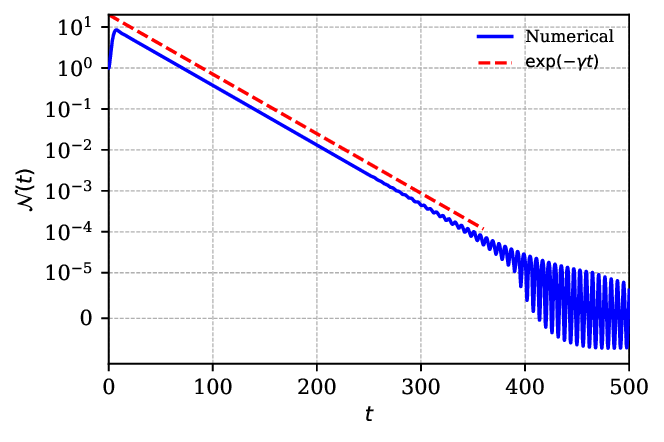}
   \caption{\footnotesize Evolution of total vorticity across the flow, ${\cal N}(t)$, shown in symlog format. Parameters: $k=1.05$, $\mu=10^{-4}$, $\gamma\equiv k\,\sigma \approx 0.0335$, $N_{\rm max}=1000$. The oscillation  period $T_{\rm t}$ at $t\gtrsim 350$ is equal to 5.98.
}
\label{fig:Evol_N}
\end{figure}

\begin{figure*} [htbp]
\centering
  \includegraphics[width=150mm]{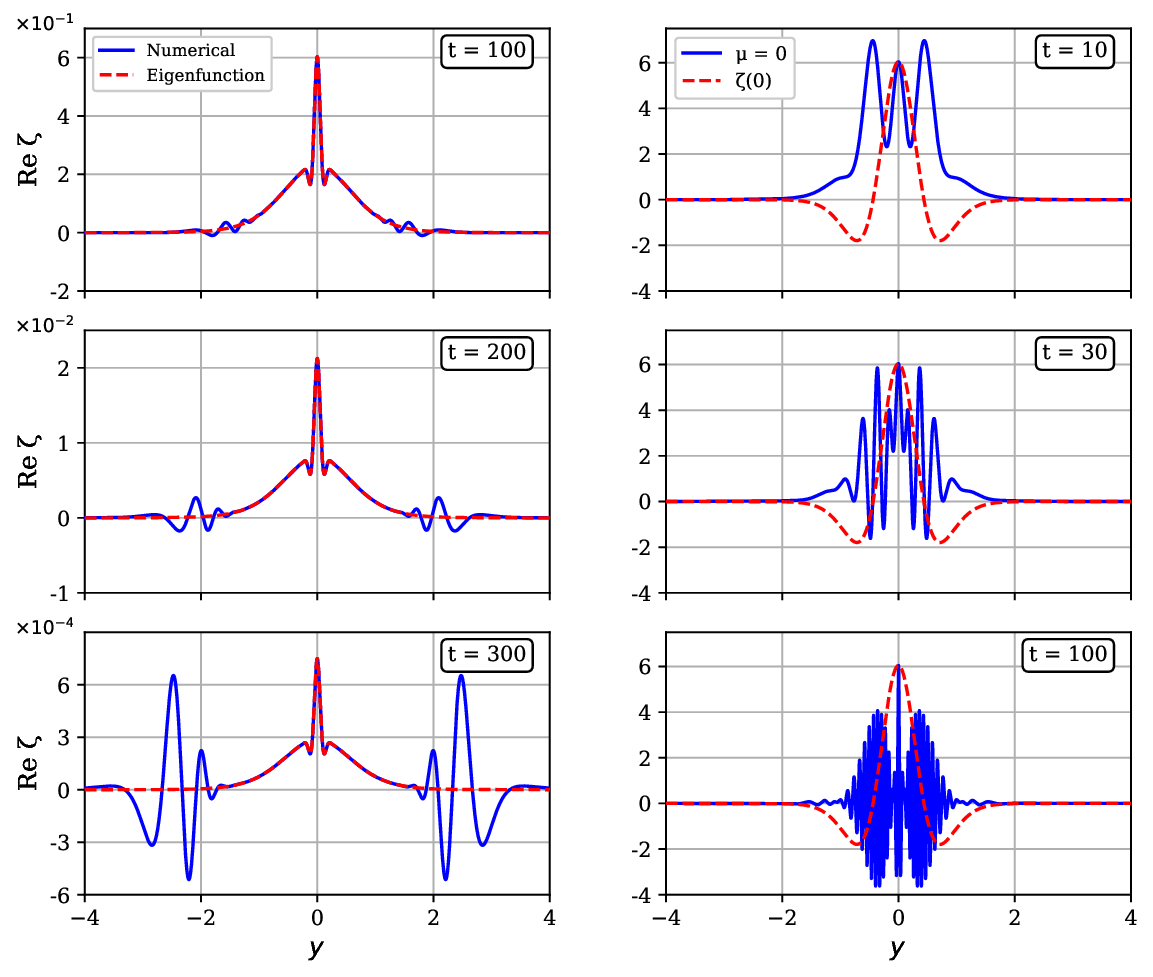}
   \caption{\footnotesize Left column: Real part of vorticity disturbance at $t=100$, 200, and 300 (blue) compared with the EF \red{of the Landau mode} (red) at the same times. Right column: Vorticity shape for inviscid flow $\mu=0$ (blue solid) and initial vorticity (\ref{eq:zeta_ini}) with $K=2$ (red dotted). Parameters: $k=1.05$, $\mu=10^{-4}$, $N_{\rm max}=1000$.
}
\label{fig:zeta}
\end{figure*}

The results of calculations for the evolution of this initial disturbance are shown in Figs.\,\ref{fig:Evol_N} and \ref{fig:zeta}.
From Fig.\,\ref{fig:Evol_N}, we observe that the total vorticity ${\cal N}(t)$ decreases exponentially after a short transition period until about $t = 350$ with a damping rate of $\gamma = \sigma\, k \approx -0.0335$, corresponding to the Landau mode with $k = 1.05$ and $\mu=10^{-4}$. Subsequently, more slowly damping modes with $|c_r|=|\omega_r|/k=1$ come into play, leading to the temporal oscillations with the expected period $T_{\rm t} = 2\pi/1.05 = 5.98$. The behavior of ${\cal N}(t)$ closely resembles that observed in the purely inviscid case, where the initial period of Landau damping is also eventually replaced by power-law decay due to the superposition of van Kampen real modes.

\red{In principle, at very long times we expect to observe a single least-damped mode with damping rate $\gamma=\sigma k\approx \mu k^3$, but numerical verification of this requires extremely precise calculations, since the expected ${\cal N}(t)$ becomes vanishingly small.}

In the viscous case, determining the law of final damping is not straightforward. Intuitively, it should remain exponential but with an averaged damping rate corresponding to the slower modes. Establishing this law numerically is difficult since the eigenvalues of the slowest modes (and their number, which becomes infinite as $N_{\rm max} \to \infty$) are very sensitive to the choice of $N_{\rm max}$.

\red{Note that in Figure~\ref{fig:Evol_N}, the exponential decay stage with the Landau damping rate is preceded by a short interval where ${\cal N}(t)$ increases. Similar growth is sometimes observed in purely inviscid fluids (see, e.g., \cite{Ellingsen_Palm1975,PS_22,PS_24}). This is transient (non-modal) growth, which can occur even when all eigenmodes (for a given $k$) are exponentially damped (or neutral, as van Kampen modes in inviscid flow) with ${\rm Im}\,c(k)\le 0$.  This arises from the non-normality of the operator governing the disturbance dynamics (see, e.g., \cite{Trefethen_1993}). The mechanism for this growth is illustrated in \cite{Shukhman_2005} using a simple dynamical model. The duration of such non-modal amplification (which may be absent entirely), as well as the relative peak value, depend on the initial disturbance form and the basic velocity profile. Recent work by Jose~\cite{Jose_2024} demonstrates that stable velocity profiles exist for which specific initial disturbances can produce significant and long-lived transient growth.} 

Fig.\,\ref{fig:zeta} (left panels) shows the evolution of the vorticity disturbance shape over time; we show only the real part, ${\rm Re}\,\zeta(y, t)$, for illustration. For comparison, the EF is shown at the corresponding times. Initially, the dominant form of the disturbance coincides with the EF form. However, as it fades, `impurities' appear associated with other, slower modes. These modes eventually begin to dominate, determining the disturbance shape. The right panels show the fragmentation of the vorticity disturbance at $\mu = 0$, which increases with time. In the purely inviscid case, the vorticity disturbance amplitude does not actually decrease compared to the initial distribution (shown by the red dotted line). However, the total vorticity ${\cal N}(t)$ still decays exponentially, as seen in Fig.\,\ref{fig:Evol_N}. This situation exactly parallels the behavior of the perturbed distribution function $f(v,t)$ and the perturbed electron density $n_e(t) = \int\d v\, f(v,t)$ in collisionless electron plasma \citep{Lan_46}.

\section{Conclusion}
\label{sec:conc}

This paper continues our investigation of perturbation dynamics in nearly inviscid shear flows, extending our understanding of how arbitrarily small viscosity transforms the eigenmode spectrum from singular van Kampen modes to regular EFs.  \red{In particular, we have illustrated the known fact that Landau quasi-modes are not true eigenmodes in inviscid flows.} Although the total vorticity ${\cal N}(t) = \int\mathrm{d}y\, \zeta(y,t)$ decays exponentially, the vorticity profiles themselves become progressively more corrugated in space over time rather than decaying self-similarly.

In inviscid shear flows, the eigenmode spectrum in the stable wavenumber region $k > k_{\rm crit}$ consists entirely of a continuous set
of singular van Kampen modes with real frequencies. We showed in our earlier work that Landau-damped solutions are, in fact, superpositions of these van Kampen modes. However, considering perturbations on complex spatial contours in the lower half-plane, we can present true EFs even for damped Landau solutions, although such contours must pass below the corresponding eigenvalue $c_L$.

Building on the insights from plasma physics studies by Ng and coauthors, we have shown that introducing viscosity, even at arbitrarily small levels, eliminates van Kampen modes and transforms Landau quasi-modes into true eigenmodes. This transformation is physically intuitive, since even weak viscosity suppresses the formation of steep vorticity gradients in space, from which van Kampen modes suffer. Our numerical evidence confirms that as the dimensionless viscosity parameter $\mu \equiv (k\, \text{Re})^{-1}$ tends to zero, the eigenvalue of the viscous problem smoothly approaches the Landau-damped solution of the inviscid problem.

Using the framework of matched asymptotic expansions with the scaling parameter $\delta = \mu^{1/3}$, we derived analytical expressions for the EF structure in the resonance region, capturing the increasingly oscillatory behavior as viscosity approaches zero. We observe numerically that the resonance region's half-width is mainly determined by Landau's damping rate $\gamma_{\rm L}$ and scales approximately as $\Delta y \approx 1.63\,\sigma$, where $\sigma \approx \sigma_{\rm L}=\gamma_{\rm L}/k$.

The peak relative amplitude ${\mathfrak P}$ of the resonance region grows without limit as $\mu \to 0$. The oscillation period $T_y$ within the resonance region is given by relation (\ref{eq:per}), demonstrating the fine-scale structure that emerges in the nearly inviscid limit. Note that we observe exactly the same structure of the EF in the distribution function $f(v)$ and its transformation as the collision frequency $\nu$ approaches zero in both plasma and \blue{homogeneous} stellar systems \citep[][]{PS_2025,PS_25PRE}.

\medskip

Our high-precision numerical calculations, employing matrix dimensions up to $N_{\max} = 2 \times 10^6$ with 1024-bit precision, validated our analytical findings and enabled us to approach the inviscid limit $\mu \to 0$ by reducing the viscosity to values as small as $10^{-9}$. The first-order viscous correction to the Landau damping rate is given by relation (\ref{eq:Delta_sigma}), with explicit expressions derived in Sec.\,\ref{subsec:2a}.

Unlike previous studies that focused primarily on stability boundaries, our approach captures the complete transformation of EF structure across the transition from stable discrete modes to oscillatory resonance patterns. The EF in the inner region takes the approximate form given by expression (\ref{eq:zeta_inner}), where the resonance structure becomes increasingly pronounced as viscosity decreases.

While the Landau mode dominates the intermediate asymptotic behavior, we also identified modes with slower damping rates from the spectrum edges that emerge only after significant delay. Our calculations demonstrate that exponential decay is clearly established, and the spatial structure $\zeta(y,t)$ exhibits distinct features of the Landau mode eigenfunction even when these slower modes contribute noticeably to the total evolution.

\red{The duration of the time interval during which the vorticity profile matches the Landau mode eigenfunction may depend on the initial perturbation shape, and less stable modes may become significant earlier, potentially eliminating this coincidence period entirely. Calculations with initial conditions similar to those presented here show that the coincidence stage persists; however, the role of the initial perturbation requires further investigation.}

The smooth transition of eigenvalues and dramatic EF transformation as viscosity vanishes reveals striking similarities to the behavior observed in \blue{homogeneous} stellar and plasma systems. Despite the vastly different mathematical frameworks -- vorticity evolution in fluid dynamics versus kinetic distribution functions -- both systems exhibit the same key features: small viscosity (or collisions) transforms quasi-modes into regular true eigenmodes with oscillatory structure in the resonance region, while eigenvalues receive only small corrections. This correspondence indicates similar mechanisms governing quasi-mode regularization in both contexts, \blue{despite significant differences between these environments in other respects.}

\blue{Unlike plasma and homogeneous stellar media, where the least damped Landau mode determines the final asymptotic decay, in shear flow the situation differs fundamentally. In the inviscid case, the flow has one or several Landau quasi-modes (eigenvalues without eigenfunctions) plus a continuous spectrum of van Kampen modes. When weak viscosity is added, the Landau quasi-modes transform into true modes with eigenfunctions, while their eigenvalues shift only slightly. The van Kampen modes disappear, replaced by a new continuous spectrum extending vertically downward from $c=\pm 1-\i\,\mu k^2$ into the lower half-plane $\text{Im}\,c<0$. This new spectrum includes modes with damping rates slower than the Landau mode. Consequently, the Landau mode governs only the intermediate-time behavior of the total vorticity ${\cal N}(t)$ (except when the initial perturbation coincides with its eigenfunction), while modes from the continuous spectrum with slower damping determine the final asymptotics.}

\medskip
\blue{Finally, the question of what conditions the velocity profile $U(y)$ must satisfy for at least one quasi-mode to exist remains unresolved. Even more interesting is the question of quasi-mode survival or non-survival (if they exist at all) when viscosity is included. As mentioned in the Introduction, there are examples of flow profiles where quasi-modes of the inviscid problem disappear when viscosity is included. These issues require separate study.}

\section*{Data Availability Statement}
The data that support the findings of this study are available in Ref.~\cite{polyachenko2025data}.
\smallskip

\begin{acknowledgments}
\vspace{-10pt}
We thank M. Weinberg for pointing out the papers by Ng, Bhattacharjee \& Skiff, and both referees for their careful reading and suggestions that improved the clarity and flow of the presentation. This research was partially supported by NSF grant PHY-2309135 to the Kavli Institute for Theoretical Physics (E. Polyachenko), and by the Russian Academy of Sciences Program No. 28 (subprogram II, `Astrophysical Objects as Cosmic Laboratories') and the Ministry of Science and Higher Education of the Russian Federation (I. Shukhman). High-precision calculations were performed with the Advanpix Multi-precision Computing Toolbox and the GNU Multiple Precision Floating-Point Reliable Library.
\newpage
\end{acknowledgments}


\bibliography{main}{}

\newpage
\begin{widetext}
\appendix

\section{Derivation of the matrix equation}\label{appA}
We write Eq.~(\ref{eq:OZ_P_n}) in the form
\begin{multline}
    c\,\sum\limits_n A_n\,\{[-n(n+1)-k^2]\,P_n +n(n+1)\,(u^2\,P_n)\}\\
=\sum\limits_n A_n \{[-(n-1)(n+2)-k^2]\, (u\,P_n)+(n-1)(n+2)\,(u^3\,P_n)\}  \\
+\i\mu \left[(1-u^2)\,{\cal L}-k^2\right]\sum\limits_n A_n \{[-n(n+1)-k^2]\,P_n+n(n+1)\,(u^2\,P_n)\}.
\label{eq:OS_uu}
\end{multline}
For brevity, we write this as
\[
c\,\Sigma_1=\Sigma_2+\i\mu\,\Sigma_3.
\]
Now we rewrite the three combinations $u\,P_n(u)$, $u^2 P_n(u)$, and $u^3 P_n(u)$ in the sums $\Sigma_{1,2,3}$ using Eq.~(8.914.1) from \cite{Gradshteyn_8}:
\be
u\,P_n=\frac{n}{2n+1}\,P_{n-1}+\frac{n+1}{2n+1}\,P_{n+1}.
\ee
We write this equality as
\be
u\,P_n=S^{(1)}_{n,-1}\,P_{n-1}+S^{(1)}_{n,+1}\,P_{n+1},
\label{eq:S1}
\ee
where
\be
S^{(1)}_{n,-1}=\frac{n}{2n+1},\ \ \ \ S^{(1)}_{n,+1}=\frac{n+1}{2n+1}.
\ee

Multiplying (\ref{eq:S1}) by $u$, we find
\be
u^2\,P_n=S^{(2)}_{n,-2}\,P_{n-2}+S^{(2)}_{n,0}\,P_n+S^{(2)}_{n,+2}\,P_{n+2},
\ee
\begin{align}
    S^{(2)}_{n,-2}&=\ S^{(1)}_{n,-1}\, S^{(1)}_{n-1,-1},\\
    S^{(2)}_{n,0\phantom{-}}&=\ S^{(1)}_{n,-1}\, S^{(1)}_{n-1,+1}+S^{(1)}_{n,+1}\,S^{(1)}_{n+1,-1}, \\
    S^{(2)}_{n,+2}&=\ S^{(1)}_{n,+1}\, S^{(1)}_{n+1,+1}.
\end{align}
Likewise
\be
u^3\,P_n=S^{(3)}_{n,-3}P_{n-3}+S^{(3)}_{n,-1}P_{n-1}+S^{(3)}_{n,+1}P_{n+1}+S^{(3)}_{n,+3}P_{n+3},
\ee
\begin{align}
S^{(3)}_{n,-3}&=S^{(2)}_{n,-2}\,S^{(1)}_{n-2,-1},\\
S^{(3)}_{n, -1}&=S^{(2)}_{n,-2}\,S^{(1)}_{n-2,+1}+S^{(2)}_{n,0}\,S^{(1)}_{n,-1},\\
S^{(3)}_{n,+1}&=S^{(2)}_{n,0}\,S^{(1)}_{n,+1}+S^{(2)}_{n,+2}\,S^{(1)}_{n+2,-1},\\
S^{(3)}_{n,+3}&=S^{(2)}_{n,+2}\,S^{(1)}_{n+2,+1}.
\end{align}

Next, we work with each of the three sums separately.

\vspace{18pt}
\centerline{\it Sum $\Sigma_1$}

\vspace{-5pt}
\be
\Sigma_1=\sum\limits_n A_n\,\Bigl\{[-n(n+1)-k^2]\,P_n +n(n+1)\,(u^2\,P_n)\Bigr\}=
\sum\limits_n A_n\,\Bigl(R_{n,-2}P_{n-2}+R_{n,0}P_n+R_{n,+2}P_{n+2}\Bigr),
\ee
\begin{align}
R_{n,-2}&=\ n(n+1)\,S^{(2)}_{n,-2},
\label{eq:R1}\\
R_{n,0\phantom{-}}&=\ n(n+1)\Bigl(S^{(2)}_{n,0}-1\Bigr)-k^2,
\label{eq:R2}\\
R_{n,+2}&=\ n(n+1)\,S^{(2)}_{n,+2}.
\label{eq:R3}
\end{align}

\vspace{10pt}
\centerline{\it Sum $\Sigma_2$}
\vspace{-5pt}
\begin{multline}
    \Sigma_2=\sum\limits_n A_n \Bigl\{[-(n-1)(n+2)-k^2]\, (u\,P_n)+(n-1)(n+2)\,(u^3\,P_n)\Bigr\}\\=
\sum\limits_n A_n\,\left[Q_{n,-3}P_{n-3}+Q_{n,-1}P_{n-1}+Q_{n,+1}P_{n+1}+Q_{n,+3}P_{n+3}\right],
\end{multline}

\begin{align}
Q_{n,-3}&=(n-1)(n+2)\,S^{(3)}_{n,-3},\\
Q_{n,-1}&=(n-1)(n+2)\Bigl(S^{(3)}_{n,-1}-S^{(1)}_{n,-1}\Bigr)-k^2 S^{(1)}_{n,-1},\\
Q_{n,+1}&=(n-1)(n+2)\Bigl(S^{(3)}_{n,+1}-S^{(1)}_{n,+1}\Bigr)-k^2 S^{(1)}_{n,+1}, \\
Q_{n,+3}&=(n-1)(n+2)\,S^{(3)}_{n,+3}.    \end{align}

\vspace{12pt}
\centerline{\it Sum $\Sigma_3$}

\vspace{-5pt}
\be
\Sigma_3=\sum\limits_n A_n\,\Bigl(T_{n,-4}P_{n-4}+T_{n,-2}P_{n-2}+T_{n,0}P_{n}+T_{n,+2}P_{n+2}+
T_{n,+4}P_{n+4}\Bigr).
\ee
Here
\begin{align}
    T_{n,-4}&=(n-2)(n-1)\,S^{(2)}_{n-2,-2}\,R_{n,-2},\\
T_{n,-2}&=\Bigl[(n-2)(n-1)(S^{(2)}_{n-2,0}-1)-k^2\Bigr]\,R_{n,-2}+n(n+1)\,S^{(2)}_{n,-2}\,R_{n,0},\\
T_{n,0}\phantom{-}&=(n-2)(n-1)\,S^{(2)}_{n-2,+2} R_{n,-2}+[n(n+1) (S^{(2)}_{n,0}-1)-k^2]\,R_{n,0}+(n+2)(n+3)\,S^{(2)}_{n+2,-2}R_{n,+2},\\
T_{n,+2}&=n(n+1)\,S^{(2)}_{n,+2}\,R_{n,0}+\Bigl[(n+2)(n+3)\,(S^{(2)}_{n+2,0}-1)-k^2\Bigr]\,R_{n,+2},\\
T_{n,+4}&=(n+2)(n+3)\,S^{(2)}_{n+2,+2}\,R_{n,+2}.
\end{align}
Thus, we have
\begin{multline}
    c\,\sum\limits_n A_n\,\Bigl(R_{n,-2}P_{n-2}+R_{n,0}P_n+R_{n,+2}P_{n+2}\Bigr)\\
=\sum\limits_n A_n\,\left(Q_{n,-3}P_{n-3}+Q_{n,-1}P_{n-1}+Q_{n,+1}P_{n+1}+Q_{n,+3}P_{n+3}\right)\\+
\i\mu\,\sum\limits_n A_n\,\Bigl(T_{n,-4}P_{n-4}+T_{n,-2}P_{n-2}+T_{n,0}P_{n}+T_{n,+2}P_{n+2}+
T_{n,+4}P_{n+4}\Bigr),
\end{multline}
where the coefficients $R_{n, m=\pm 2,0}$, $Q_{n,m=\pm 3, \pm 1}$, $T_{n,m=\pm 4, \pm 2, 0}$ are given by the formulas given above. Equating the coefficients for the same $P_n$ yields:
\begin{multline}
    c\,\Bigl(R_{n-2,+2}\,A_{n-2}+R_{n,0}\,A_n+R_{n+2,-2}\,A_{n+2}\Bigr)\\ 
=\left(Q_{n-3,+3}A_{n-3}+Q_{n-1,+1}A_{n-1}+Q_{n+1,-1}A_{n+1}+Q_{n+3,-3}A_{n+3}\right)\\
+\i\mu\,\Bigl(T_{n-4,+4}A_{n-4}\!+\!T_{n-2,+2}A_{n-2} +T_{n,0}A_{n}+
T_{n+2,-2}A_{n+2}\!+\!T_{n+4,-4}A_{n+4}\Bigr).
\label{eq:A1}
\end{multline}
This equation can be represented in the form 
 \be
 c\,\sum_{m=0}^\infty {\cal R}_{nm}\,A_m =\sum_{m=0}^\infty{\cal Q}_{nm}\,\,A_m+\i\mu\,\sum_{m=0}^\infty{\cal T}_{nm}\,\,A_m,
 \label{eq: matrix2}
 \ee
where
\be
{\cal R}_{n,n}=R_{n,0},\ \ \ {\cal R}_{n,n\mp 2}=R_{n\mp 2,\pm 2},
\ee
\be
{\cal Q}_{n,n\mp 1}=Q_{n\pm 1,\pm 1},\ \ \ {\cal Q}_{n,n\mp 3}=Q_{n\pm 1,\pm 3},
\ee
\be
{\cal T}_{n,n}\!=\!T_{n,0},\ \ {\cal T}_{n,n\mp 2}\!=\!T_{n\mp 2,\pm 2}, \ \ {\cal T}_{n,n\mp 4}\!=\!T_{n\mp 4,\pm 4},
\ee
where ${\cal R}_{nm}$ is a matrix with 3 nonzero diagonals, $m=n, n\pm 2$, ${\cal Q}_{nm}$ is a matrix with 4 nonzero diagonals, $m=n\pm 1, n\pm 3$, and ${\cal T}_{nm}$ is a matrix with 5 nonzero diagonals, $m=n, n\pm 2, n\pm 4$.

\section{Continuous spectrum}\label{appB}

We explain the origin of the continuous spectrum for the free flow $U=\tanh y$ in viscous case. 
The asymptotic form of Eq. (\ref{eq:OZ}) as $y\to\pm \infty$ is a fourth-order differential equation with constant coefficients
\be
(c\mp 1)({\d^2}/{\d y^2}-k^2)\psi = \i\,\mu\,({\d^2}/{\d y^2}-k^2)^2\psi.
\ee
Seeking a solution of the form  $\psi\sim \exp(\lambda y)$ yields
\be
-(c\mp 1)(k^2-\lambda^2)=\i\mu\,(k^2-\lambda^2)^2,
\ee
which give s  four values of $\lambda$ at each infinity:
\be
\lambda_{1}= -k, \quad \lambda_2=k,
\ \lambda_{3}=-\sqrt{k^2-\i\,(c\mp 1)/{\mu}}, \ \ \ \ \
   \lambda_{4}=\sqrt{k^2-\i\,(c\mp 1)/{\mu}},
   \label{eq:lambda}
  \ee
where the upper sign corresponds to  $y\to\infty$ and the lower sign to $y\to -\infty$. The square root $\sqrt{k^2-i\,(c\mp 1)/{\mu}}$ denotes the branch with positive real part, giving ${\rm Re}\, \lambda_3<0$ for $y\to\infty$ and ${\rm Re}\, \lambda_4>0$ for $y\to -\infty$. These $\lambda$ were first derived by Betchov \& Szewczyk \cite{Betchov_Szewszyk_1963}.

Continuous spectrum follows directly from this analysis if we replace the boundary conditions ($\psi$ and $d\psi/dy$ vanishing at infinity) with boundedness condition. This  requires that   $\lambda_3$ and $\lambda_4$ be purely imaginary  as $y\to \infty$ or $y\to-\infty$, which yields
\be
c_r=1,\quad {\rm or}\quad c_r=-1,\quad {\rm and}\quad c_i/\mu+k^2=-k^2 p^2 \le 0,
\ee
The continuous spectrum therefore consists of two vertical lines:
 \be
 c=\pm 1-\i\mu k^2(1+p^2),\quad 0\le p<\infty.
 \ee

\end{widetext}
\end{document}